\documentclass[useAMS,usenatbib,rotating]{mn2e}

\pdfoutput=1
\usepackage{deluxetable}
\usepackage{amssymb}
\usepackage{natbib}
\usepackage{multirow}
\usepackage{hyperref,color}
\usepackage{epsfig}
\usepackage{rotating}
\usepackage{lscape}

\title[Galaxies' SFHs from $V,I,K^\prime$ photometry of Globular Clusters]{New insights into the star formation histories of candidate intermediate-age early-type galaxies from $K^\prime$-band imaging of globular clusters}
\author[I.\,Y.\,Georgiev et al.]{
Iskren Y. Georgiev$^{1}$\thanks{E-mail:iskren@astro.uni-bonn.de},
Paul Goudfrooij$^{2}$
and
Thomas H. Puzia$^3$
\\
$^1$Argelander Institut f\"ur Astronomie der Universit\"at Bonn, Auf dem H\"ugel 71, D-53121 Bonn, Germany\\
$^2$Space Telescope Science Institute, 3700 San Martin Drive, Baltimore, MD 21218, USA\\
$^3$Department\,of\,Astronomy and Astrophysics,\,Pontificia Universidad Católica de Chile,\,Av.\,Vicuna Mackenna\,4860,\,7820436\,Macul,\,Santiago,\,Chile
}
       
\begin{document}

\date{Accepted 2011 November 1}

\pagerange{\pageref{firstpage}--\pageref{lastpage}} \pubyear{2011}

\maketitle

\label{firstpage}

\begin{abstract}
We investigate the age and metallicity distributions of bright globular 
clusters (GCs) in the candidate intermediate-age early-type galaxies 
NGC\,3610, NGC\,584 and NGC\,3377 using a combination of new Gemini/NIRI 
$K^\prime$-band imaging and existing optical $V,I$ photometry from Hubble 
Space Telescope data. The $V\!-\!I$ versus $I\!-\!K'$ colour-colour diagram is found 
to break the age-metallicity degeneracy present in optical colours and 
spectroscopy, as $I\!-\!K^\prime$ primarily measures a populations' metallicity. In 
addition, it is relatively insensitive to the effect of hot horizontal 
branch (HB) stars that are known to be present in massive old GCs. By 
interpolation between Simple Stellar Population model tracks we derive 
photometric cluster ages, metallicities and masses. In general we find that 
"metal-poor" GCs with $[Z/\mbox{H}] \la -0.7$\,dex are older than more metal-rich 
GCs. For the most massive GCs (${\cal{M}}\geq6\times10^5M_\odot$) in NGC\,3610 
with available spectroscopic data, the photometric ages are older by $\sim2$\,Gyr, 
and this difference is more pronounced for the metal-poor GCs. However, the 
photometric and spectroscopic metallicities are in good agreement. We suggest 
that this indicates the presence of a hot HB in these massive clusters, which 
renders spectroscopic ages from Balmer line strengths to be underestimated. 
To support this suggestion we show that \emph{all} Galactic GCs with 
${\cal{M}}\geq6\times10^5M_\odot$ feature hot HBs, except 47\,Tuc. Using a 
recent observational relation between the luminosity of the most massive GC 
and the galaxy's SFR at a given age, we find that the galaxies' peak SFR was 
attained at the epoch of the formation of the oldest (metal-poor) GCs. The 
age and [$Z$/H] distributions of the metal-rich GCs are broad, indicating 
prolonged galaxy star formation histories. The peak ages of the metal-rich GCs 
in the sample galaxies are 3.7, 5.9, and 8.9 Gyr for NGC\,3610, NGC\,584, and 
NGC\,3377 respectively. The peak value of the age and metallicity distributions 
of the GCs is correlated with the host galaxies' luminosity-weighted age and 
metallicity, respectively, indicating that the GCs can indeed be used as 
relevant proxies of the star formation histories of galaxies.
\end{abstract}


\section{Introduction}

Globular clusters (GCs) are among the very few observable fossil records 
of the formation and evolution of galaxies. The nature of their integrated 
properties, closely approximated by a Simple Stellar Population (hereafter SSP),
significantly simplifies the determination of their ages and metallicities
relative to that of the diffuse light of their parent galaxies. Hence they
contain crucial information about the history of star formation and chemical
enrichment of their parent galaxies \cite[e.g.,][]{Whitmore99, Goudfrooij01b}. 
A key development in this respect has been the discovery that the GC systems 
of many early-type galaxies have bimodal colour distributions \cite[e.g.,][]
{Kundu&Whitmore01, Larsen01, Peng06}, providing clear evidence for the 
occurrence of a "second event" in the formation of these systems. Such a 
bimodal distribution was actually predicted from merger scenarios of E galaxy 
formation by \cite{Schweizer87,Ashman&Zepf92}, who suggested that young GCs 
form out of chemically enriched gas during major gas-rich mergers, then redden 
and fade with time to produce the "red" population of GCs seen in old 
ellipticals. Since this prediction, populations of young and intermediate-age 
GCs have indeed been found in several gas-rich merger remnants (e.g. 
\citeauthor{Holtzman92} \citeyear{Holtzman92}; \citeauthor{Schweizer96} \citeyear{Schweizer96}; 
\citeauthor{Miller97} \citeyear{Miller97}; \citeauthor{Whitmore99} \citeyear{Whitmore99}; 
\citeauthor{Goudfrooij01} \citeyear{Goudfrooij01} \citeyear{Goudfrooij01b}b; 
\citeauthor{Goudfrooij07} \citeyear{Goudfrooij07}). 

Although HST studies have provided much information on the bimodality of 
GC systems, the typically used $V\!-\!I$ or $g-z$ colours provide only limited 
constraints on the age and metallicity of GCs. This is due to the well-known 
age-metallicity degeneracy, where increasing age or metal content have a 
similar effect in reddening the optical colours. The non linear relation 
between optical colour and metallicity adds an additional complication in 
the interpretation of colour bimodality \citep{Yoon06}. In principle, 
spectroscopy is a more accurate way to estimate age and metallicity. However, 
in practice this is very time consuming even for an 8-10m class telescope, 
and only 1-2 handfuls of GCs per galaxy have been studied in this way to 
date with sufficient precision \cite[e.g.,][]{Forbes01,Goudfrooij01,
Puzia05,Woodley10,Alves-Brito11}. 

Another way to break the 
age-metallicity degeneracy is to combine optical and near-infrared colours. 
In the optical, hot evolved stars (e.g., horizontal branch (HB) stars and 
blue straggles) and young main sequence stars (if present) contribute 
significantly to the integrated light. They also enhance the strength of 
high-order Balmer lines, which causes a bias toward younger ages when 
classically derived from such absorption lines (see more in Sect.\,\ref{Subsect:Photometric Age Z}). 
Conversely, for ages $> 1$ Gyr, the near-IR is completely dominated by 
light from red giant stars. Thus near-IR colours like $I\!-\!K$ or $J-K$ measure 
the typical temperature of the giant branch, which is known to be directly 
sensitive to metallicity without any significant age dependence \cite[e.g.,][]{Worthey94}. 
Relative to optical spectroscopy, the "optical+near-IR" imaging technique 
has the additional advantage that a larger number GCs can be detected in 
a reasonable observing time per galaxy. This technique was pioneered for 
GCs in intermediate-age galaxies by 
\cite{Puzia02} \cite[see also][]{Goudfrooij01} using ground-based (ESO/VLT) $K$-band imaging in combination 
with Wide Field Planetary Camera 2 (WFPC2) $V$- and $I$-band data from {\it 
HST} to study NGC 3115 and NGC 4365, two nearby early-type galaxies. 
\citet{Puzia02} found that NGC 4365 hosts a significant number of 
intermediate-age ($2-6$ Gyr old) GCs, whereas NGC 3115 only hosts "old" GCs. 
The age and metallicity values found for GCs by \citet{Puzia02} were confirmed
by subsequent spectroscopy of a handful of GCs in these two galaxies 
\citep{Kuntschner02,Larsen03}. Finally, the complications with the non-linear 
relation between optical colour and metallicity is safely avoided with 
optical-NIR colours \cite[e.g.][]{Cantiello&Blakeslee07,Kundu&Zepf07,Spitler08b}.
Hence, the optical+near-IR imaging technique is a valuable alternative to
spectroscopy in studying GC age and metallicity distributions \cite[see
also][]{Chies-Santos11}. 

The discovery of a significant number of intermediate-age GCs in
a fairly "normal" elliptical galaxies like NGC 4365 (\citeauthor{Puzia02} 
\citeyear{Puzia02}, \citeauthor{Kundu05} \citeyear{Kundu05}, but see 
\citeauthor{Brodie05} \citeyear{Brodie05}, \citeauthor{Chies-Santos11} \citeyear{Chies-Santos11})
and NGC\,5813 (\citeauthor{Hempel07} \citeyear{Hempel07}a, \citeyear{Hempel07b}b) 
highlights the power of GCs in probing major star formation events in galaxies. 
While such an intermediate-age population in NGC 4365 might not seem
unexpected given the presence of a kinematically decoupled core, it has not been
identified from the galaxy's integrated light. This raises the important
question whether or not early-type galaxies, which show signs of past
interactions or mergers, generally host intermediate-age, metal-rich  
subpopulations. 

In this paper we use the optical/near-IR technique mentioned above
to constrain the presence of intermediate-age population(s) of
GCs in a small sample of nearby early-type galaxies that show signs of having
undergone a dissipative merger several Gyr ago. Our target galaxies 
are NGC\,3610, NGC\,3377 and NGC\,584 described further in the next section.

\section{The Target Galaxies} \label{Sect:Targets}

The target galaxies were selected to have the following properties at the time
of proposal submission: 
{\it (i)} The galaxy is an E or E/S0 galaxy that is nearby enough
($v_{\rm hel}<2000$\,km\,s$^{-1}$) to detect several tens of GCs in
the $K$ band in reasonable exposure times, {\it (ii)} the galaxy has a
rich GC system with \emph{high-quality} optical GC colours measured
with {\it HST} \citep[$V\!-\!I$;][]{Kundu&Whitmore01}  
whose distribution does not show a clear bimodality \cite[the absence 
of such bimodality is expected in the presence of a significant subpopulation 
of GCs that is metal rich and of intermediate-age; see also][]{Kissler-Patig98}, 
and {\it (iii)\/} measurement of (luminosity-weighted) age from 
integrated-light spectroscopy or photometry is in the range 2\,--\,5 Gyr. 
Table~\ref{tab:sample} lists relevant basic properties of the galaxies in 
our sample. Population properties of the sample galaxies from recent 
literature is discussed in more detail in Sect.\,\ref{Sect:GC vs galaxies Ages and Z}). 
\begin{table*}
\begin{center} 
\caption{Basic properties of the sample galaxies.
 \label{tab:sample}}
\begin{tabular}{@{\extracolsep{\fill}}lllccccccccc@{}}    
\multicolumn{3}{c}{~~} \\ [-2.5ex]
    \hline \hline
\multicolumn{3}{c}{~~} \\ [-2.2ex]
\multicolumn{1}{c}{Galaxy} & \multicolumn{2}{c}{Classification} & $A_V$ &
 \multicolumn{1}{c}{$(m-M)_0$} & D &  \multicolumn{1}{c}{$M_V$} & $(V\!-\!I)_0$ & $(I\!-\!K)_0$ & Age$_{\rm spec}$  & [Z/H]$_{\rm spec}$ & SFR$_{\rm Peak}$\\ [0.5ex]  
 \ & RSA & RC3 & [Mag] & [Mag] & [Mpc] & [Mag] & [Mag] & [Mag] & [Gyr] & [dex] & [$M_\odot/$yr]\\ [0.5ex]
\multicolumn{1}{c}{(1)} & (2) & (3) & (4) & (5) & (6) & (7) & (8) & (9) & (10) & (11) & (12) \\ [0.5ex] \hline 
\multicolumn{3}{c}{~~} \\ [-0.8ex]                                                
NGC 3610 & E5    & E5: & 0.03 & $32.71 \pm 0.08$\tablenotemark{c} & 34.84 & $-21.99$ & 1.14 & 1.96 & $1.7 \pm 0.1$\tablenotemark{e} & $+0.76 \pm 0.16$\tablenotemark{e} & 234\\ [0.5ex]
NGC 584  & S0$_1$ & E4 & 0.14 & $31.52 \pm 0.20$\tablenotemark{a} & 20.14 & $-21.31$ & 1.14 & 1.91 & $2.6 \pm 0.3$\tablenotemark{d} & $+0.46 \pm 0.03$\tablenotemark{d} & 101\\
NGC 3377 & E5-6   & E6 & 0.11 & $30.17 \pm 0.16$\tablenotemark{b} & 13.87 & $-20.01$ & 1.11 & 1.81 & $3.7 \pm 0.9$\tablenotemark{d} & $+0.19 \pm 0.05$\tablenotemark{d} & 75\\
\hline
\end{tabular}\vspace{-1cm}
\tablecomments{Column descriptions: (1): Galaxy name. (2) and (3):
  Classification in RSA \citep{RSA81} and RC3 \citep{RC3} catalogs,
  respectively. (4): Foreground reddening in $A_V$ according to
  \citep{Schlegel98}. (5): Distance Modulus according to reference given (see
  below) and the corresponding distance in Mpc in (6) and galaxy absolute magnitude in (7). (8) and (9): $(V\!-\!I)_0$ and $(I\!-\!K)_0$ colours according to \citet{Michard05}. (10) and (11): Spectroscopic age and [Z/H] according to reference given (see below). (12): Galaxy peak SFR calculated in this study (see Sect.\,\ref{Sect:Gal_SFR}).\\
$^a$\cite{Tonry01}; $^b$\cite{Harris07}; $^c$\cite{Cantiello07}; $^d$\cite{Trager00}; $^e$\cite{Howell05}.
} 
\end{center}
\end{table*}

\section{Observations and Data Reduction}\label{Sect:Obs.n.Red}

\subsection{Observations} \label{SubSect:Observations}

We performed near-infrared $K^\prime$-band ($2.12~\mu m$) imaging of 
NGC\,3610, NGC\,3377, and NGC\,584 with the Near-Infra-Red Imager and 
Spectrometer (NIRI) instrument on Gemini North \citep{Hodapp03}. Our 
program ID was GN-2004A-Q-17. We used the f/6 NIRI camera imaging mode, 
which provides a plate scale of 0.117 arcsec/pixel and a field of view 
of $2\times2$  arcmin$^2$, similar to the field of view of the WFPC2 
and ACS cameras aboard {\it HST}.

\begin{figure*}
\epsfig{file=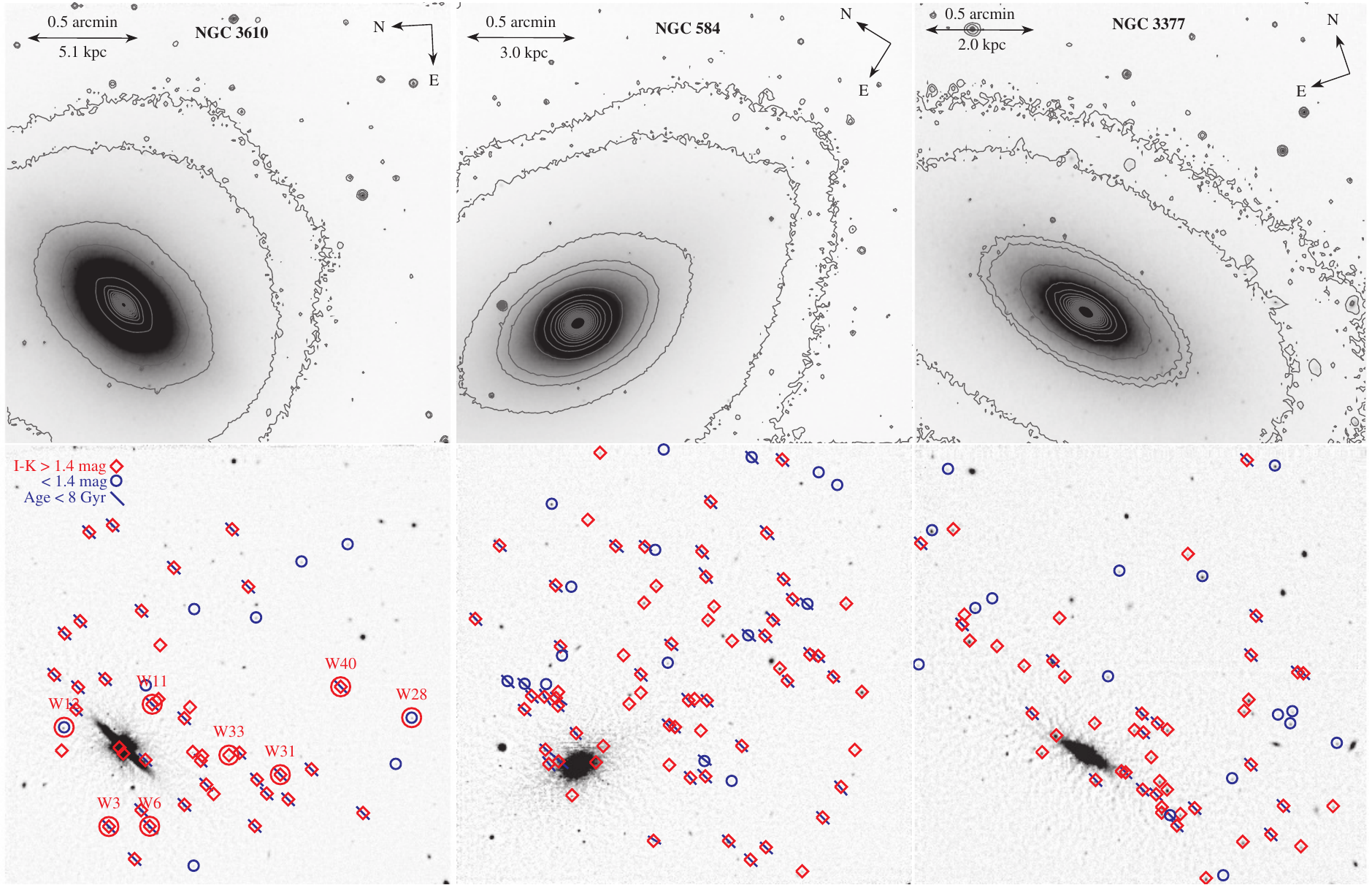, width=0.5\textwidth, bb=135 0 400 351}
\caption{$K^\prime$-band Gemini/NIRI images of NGC\,3610, NGC\,584 
and NGC\,3377 from left to right, respectively. The outer four isointensity 
contours on the top panels indicate surface brightnesses of $5,10,50$ and 
$100\times\sigma$ above the background. Bottom panels show the galaxies with 
median filtered images subtracted. Different symbol types indicate globular 
clusters bluer (circles) or redder (diamonds) than $I\!-\!K=1.4$\,mag or younger 
than 8\,Gyr (tilted solid line). NGC\,3610 GCs with spectroscopic measurements 
in the literature are labeled and shown with larger circles.
}\label{fig:N3610_N584_N3377_mos2}
\end{figure*}

In order to reliably detect the presence of both "blue" and "red" GCs, 
with S/N$\sim\!5$ similar to that of the WFPC2 data, the targeted limiting 
magnitude was the typical turnover magnitude of GC systems of "normal" 
galaxies: $M_V\!=\!-7.3$\,mag \citep[e.g.,][]{Kundu&Whitmore01,Larsen01}, 
corresponding to $M_K\!=\!-9.8$ for a typical colour of an "old" (age
of 14 Gyr) SSP with [Fe/H$]\!=\!-1.5$ \citep[e.g.][]{BC03}. The exposure
times to reach the limiting magnitudes (cf. Table\,\ref{Table:ObsLog}) 
were estimated according to the distance moduli and were such that 
NGC\,3377 was observed in a single night, while NGC\,3610 and NGC\,584 
required observations to be spread over two and four nights, respectively. 
Table\,\ref{Table:ObsLog} summarizes the observations and 
night conditions.
\begin{table}
\caption{Log of Observations \label{Table:ObsLog}}
\begin{tabular}{cccp{.5cm}c}
\hline \hline \\
Target & Nights & FWHM & $\rm N$ & Total Exposure \\ 
&&&& \\ [-2.5ex]
& {\footnotesize yyyy-mm-dd} & arcsec & images & seconds\\
\hline \hline
\multirow{2}{*}{NGC 3610} & 2004-02-10 & $0.43\arcsec$ & 64 & 3840 \\
& 2004-02-11 & $0.35\arcsec$ & 55 & 3300 \\
\hline
\multirow{4}{*}{NGC 584} & 2004-07-28 & $0.45\arcsec$ & 35 & 2100 \\ 
&  2004-07-29 & $0.46\arcsec$ & 34 & 2040 \\ 
&  2004-07-30 & $0.53\arcsec$ & 31 & 1860 \\ 
&  2004-07-31 & $0.41\arcsec$ & 38 & 2280 \\ 
\hline
NGC 3377 & 2004-03-02 & $0.51\arcsec$ & 28 & 1680 \\ 
\hline\hline
\end{tabular}
\end{table}
In Figure\,\ref{fig:N3610_N584_N3377_mos2} we show the spatial coverage 
of the NIRI observations. The character of the spatial distribution of detected GCs 
reflects the overlap between the NIRI and the HST/ACS,WFPC2 fields of view. Globular 
clusters shown with different symbols reflect their properties as derived in 
Section\,\ref{Sect:VIK colour-colour plot, ages and Zs}.

\subsection{Data Reduction} \label{SubSect:Reduction}

Image reduction was performed with tasks of the 
{\sc gemini/niri}\footnote{http://www.gemini.edu/sciops/data-and-results/processing-software} 
packages within {\sc iraf}\footnote{IRAF is distributed by the National 
Optical Astronomy Observatories, which are operated by the Association of 
Universities for Research in Astronomy, Inc., under cooperative agreement 
with the National Science Foundation.}. Due to the "first 
frame"\footnote{First frame problem arises because the NIRI array is not 
continuously reset when idle. Thus, each new exposure after idle has 
different dark current level, probably due to image persistence after 
saturating the array (NIRI has no shutter).} 
problem of the NIRI detector, all first exposures (dark, flat and on-/off-target 
images) of each new sequence were excluded before basic image processing. 

Normalized flats were constructed from images taken with the 
calibration unit shutter closed ("lamps off") subtracted from exposures 
taken with the shutter open ("lamps on") and using short darks to identify 
bad pixels. Flat field images with the IR lamp on and off allow separation 
of the instrumental thermal signature from the sensitivity response. 

The near-IR  sky level and structure varies on time scales of a few minutes. 
To account for such variations the imaging sequence of four on- followed 
by four off-target exposures provides close enough in time blank fields 
to derive the sky flat and correct the on-target exposures. 

Some of the NIRI frames contain an electronic pattern visible as vertical 
striping with a period of eight columns and not always present in all 
quadrants. Before including the affected frames in our image processing 
list we used the stand-alone python routine, {\it nirinoise.py}\footnote{http://staff.gemini.edu/$\sim$astephens/niri/patternnoise/} 
which almost perfectly removed the striping. The same image reduction 
steps were applied to the standard star observations as well. We discarded 
few images which had trails or double-peaked stellar profiles reflecting 
the loss of the guide star resulting in improper telescope guiding.

All reduced science and standard star exposures of the corresponding 
nights were registered to a common coordinate system with {\sc niri/imcoadd} 
based on {\sc geomap} geometric solutions. The final science image of 
each galaxy is derived from average combined individual images scaled 
to that of the most photometric night for the case of NGC\,3610 and 
NGC\,584, which were observed over few nights. The total number 
of combined individual exposures for NGC\,3610, NGC\,3377, and NGC\,584 
is N=119, N=28, and N=137, respectively. The result is one deep stacked 
$K^\prime$-band image per galaxy.

\section{Detection, photometry and calibration} \label{Sect:GC.Detect.Phot.Cal}

\subsection{Detection and photometry}\label{Sect:DetPhot}

Globular clusters in our ground based images would appear as unresolved 
point sources due to the relatively large NIRI pixel scale at the distances 
to the galaxies. For the nearest galaxy in our sample NGC\,3377 at a 
distance of 10.8\,Mpc we have a projected spatial resolution of 6\,pc/pix. 
This is twice the typical half-light radius ($\sim3$\,pc) of a GC 
\cite[e.g.][]{Barmby07,Masters10}. NGC\,3610 and NGC\,584 are at much 
greater distances of 34.8 and 20.1\,Mpc, respectively. Thus, even for the 
most nearby galaxy NGC\,3377, the GCs on the ground-based images will 
appear as point sources with a stellar PSF, which determined the type 
of detection and photometry of GCs.

To detect\footnote{All reduction and analysis has been performed with 
IRAF procedures.} 
as many GCs as possible to match with existing deeper HST 
photometry for these galaxies \citep{Goudfrooij07,Kundu&Whitmore01}, 
we first created images representing the integrated light component of the
galaxy by smoothing the combined images with a circular median filter with 
a radius of 8 times the seeing FWHM of the input image. The choice of the
median filter radius resulted from tests designed to render the residual
effect of point sources in the images negligible. This was done by measuring  
the magnitude difference between the original and median-subtracted images of
several isolated non-saturated point sources. The original combined image
was then divided by the square root of the smoothed image, providing an image
with uniform shot noise characteristics \citep[e.g.,][]{Goudfrooij07}. This
facilitates the uniform detection of point sources at a detection threshold of
$3\sigma$ above the background, using the {\sc daofind} task. The resulting 
object coordinates were used as input for PSF photometry performed on images 
from which the smooth galaxy light was subtracted. PSF models with radius 
3\,$\times$ the point source FWHM were created using several non-saturated 
point sources. PSF-fitting magnitudes were scaled using aperture photometry 
with a radius of 4 pix for the PSF stars. To estimate aperture corrections 
from curve-of-growth analysis, photometry of the PSF stars was performed with 
aperture radii of $r=2,4,12,17,21,31$\,pixels. We found very similar aperture 
corrections for the PSF magnitudes among the three sample galaxies ($-0.361\pm0.008$\,mag). 

\subsection{Photometric Calibration}\label{Sect:PhotCal}

Photometric calibration was performed using photometric standards (FS\,6, 
FS\,23, FS\,34, FS\,103, FS\,111, FS\,126, FS\,130, FS\,131 and FS\,134) 
observed during the same nights as the target galaxies. Their instrumental 
magnitudes were measured with the same aperture as their measured $K^\prime$ 
catalog magnitudes in \cite{Leggett06}. Thus, we derive consistent 
transformation solutions from instrumental to standard photometric 
system for all objects in our photometric lists. To allow for optical-to-near-IR 
colour terms in the photometric calibration of the $K'$-band data, we used the $V$-band 
magnitudes of the photometric standard stars in Table\,1 of \cite{Hwarden01},
who compiled measurements from \citet{Sandage&Katem82,Lasker88,McCook&Sion87,
Carlsberg89,Leggett92,Landolt92}. Hence, the derived photometric 
transformations will take into account the $V-K$ colour of the object.

We evaluated a least squares fit to the following photometric transformation 
equation:\
\begin{equation}\label{eqn:calib}
mK^\prime=K^\prime+ZP_{K^\prime}+c1\times\,X_{K^\prime}+c2\times(V-K^\prime)
\end{equation}
where $mK^\prime$ is the instrumental magnitude, $ZP_{K^\prime}$ is the 
zero-point, $c1$ and $c2$ are the airmass and colour term coefficients. 
The zero-point and airmass coefficients were fitted around their typical 
values\footnote{\protect\href{http://www.gemini.edu/?q=node/10104}{http://www.gemini.edu/?q=node/10104}} 
$ZP_{K^\prime}=23.68$\,mag, $c1=0.059$\,mag/airmass. However, a best fit 
(low rms and coefficients uncertainty) was achieved for fixed airmass 
coefficient at $c1=0.07$\,mag/airmass\footnote{The mean value found for 
a typical variation of the water vapor amount on Manua Kea \citep{Tokunaga02}}, 
while the zero-point and colour term coefficients were kept variable. 
This resulted in stable solutions for the zero-point and the $V-K$ colour 
term coefficients, close to their typical values (cf. Table\,\ref{Table:coeff}). 
Due to the only one standard star observed during the NGC\,3377 observing 
run, we fit the $ZP_{K^\prime}$ using a constant value for the colour term, which 
is the mean value obtained from the NGC\,3610 and NGC\,584 calibrations. 
Excluding both, $c1$ and $c2$ from the fit results in a $ZP_{K^\prime}=-23.73$\,mag, 
i.e. within the uncertainty estimate of the $ZP_{K^\prime}$ value with 
the airmass and colour coefficients included. All transformation coefficients 
are summarized in Table\,\ref{Table:coeff}. 
\begin{table}
\caption{Calibration Coefficients for Equation\,(\ref{eqn:calib})\label{Table:coeff}}
\begin{tabular}{l|ccc}
\hline \hline \\
& $ZP_{K^\prime}$ & $c1$\footnotemark[1] & $c2$ \\
\hline
NGC\,3610 & $-23.66\pm0.03$ & $0.07$ & $-0.025\pm0.024$\\
NGC584 & $-23.68\pm0.04$ & $0.07$ & $-0.027\pm0.021$\\
NGC\,3377 & $-23.68\pm0.03$ & $0.07$ & $-0.026$\hspace*{1.215cm}\\
\hline \hline
\end{tabular}
\footnotetext[1]{The mean value for a typical water vapor variation on Manua 
Ke \protect\citep{Tokunaga02}}
\end{table}

To convert the GCs' instrumental to standard $K^{\prime}$ magnitudes, 
$V$-band magnitudes of detected GCs in our images were adopted from the 
HST/ACS and WFPC2 optical $V,I$ photometry in \cite{Kundu&Whitmore01} for
NGC\,584 and NGC\,3377 and \cite{Goudfrooij07} for NGC\,3610. 
To match the NIRI and HST photometric lists, we transformed the ACS 
and WFPC2 GC pixel coordinates to the NIRI image pixel coordinate system 
by deriving geometric transformation solutions with {\sc geomap} and applying 
them with {\sc geoxytran} IRAF routines. The transformation from instrumental 
to $K^\prime$ magnitudes of our sample GCs was performed by evaluating 
calibration equation (\ref{eqn:calib}), taking into account the $V-K^\prime$ 
colour of the cluster. In Table\,\ref{Table:VIK_ageZM} we list the photometric properties 
of all GCs with $K^\prime$ photometry which are discussed in the following.
\begin{landscape}
\begin{table}
\centering
\caption[Magnitudes, colors, photometric age, metallicity and mass of GC]
{Magnitudes, colors, photometric age, metallicity and mass of globular 
clusters in the three post-merger remnant galaxies corrected for foreground reddening. 
Optical $V,I$ magnitudes, coordinates and clusters IDs are from \cite{Goudfrooij07} 
for NGC\,3610 and \cite{Kundu&Whitmore01} for NGC\,584 and NGC\,3377. The second 
IDs for the NGC\,3610 GCs are the IDs adopted in the spectroscopic study by 
\cite{Strader03,Strader04}. Full version of this table is available online.
}\label{Table:VIK_ageZM}
\begin{tabular}{ccccccccc}
\hline\hline
ID & RA,DEC\,(2000) & $M_V$ & $V_0$ & $(V-I)_0$ & $(I-K^\prime)_0$ & Age & [Z/H] & $M$ \\ 
& [hh:mm:ss],[dd:mm:ss] & [mag] & [mag] & [mag] & [mag] & [Gyr] & [dex] & [$10^5M_\odot$] \\
(1) & (2) & (3) & (4) & (5) & (6) & (7) & (8) & (9)\\
\hline
\multicolumn{9}{c}{\bf NGC\,3610}\\
G2/W3   & 11:18:28.12  $+$58:47:15.8 & $-11.16$ & $21.548\pm0.007$ & $  1.062\pm 0.01$ & $1.406\pm0.031$ & $ 7.896\pm^{0.785}_{2.174}$ & $-0.610\pm^{0.169}_{0.078}$ & $43.359\pm^{ 1.679}_{4.579}$\\
G4/W6   & 11:18:28.19  $+$58:47:04.9 & $-11.05$ & $21.659\pm0.007$ & $  1.054\pm 0.01$ & $1.586\pm0.026$ & $ 5.895\pm^{1.658}_{0.600}$ & $-0.452\pm^{0.051}_{0.183}$ & $33.814\pm^{ 2.819}_{1.027}$\\
G10/W11  & 11:18:23.96  $+$58:47:02.5 & $-10.73$ & $21.983\pm0.008$ & $  1.125\pm0.012$ & $1.868\pm0.024$ & $ 4.565\pm^{0.877}_{3.325}$ & $ 0.099\pm^{0.236}_{0.093}$ & $27.291\pm^{ 5.201}_{19.696}$\\
G11/W12  & 11:18:24.61  $+$58:47:26.5 & $-10.70$ & $22.011\pm0.008$ & $   0.95\pm0.012$ & $1.183\pm0.051$ & $16.210\pm^{3.309}_{8.003}$ & $-1.303\pm^{0.464}_{0.656}$ & $37.870\pm^{ 5.990}_{14.470}$\\
G26/W28  & 11:18:24.88  $+$58:45:53.0 & $ -9.87$ & $22.836\pm0.013$ & $  0.984\pm0.017$ & $1.298\pm0.062$ & $11.047\pm^{6.541}_{5.054}$ & $-0.923\pm^{0.185}_{0.533}$ & $13.858\pm^{ 5.163}_{3.987}$\\
G30/W31  & 11:18:26.62  $+$58:46:29.0 & $ -9.64$ & $ 23.07\pm0.014$ & $  1.059\pm0.019$ & $1.533\pm0.056$ & $ 6.240\pm^{1.316}_{1.011}$ & $-0.498\pm^{0.082}_{0.142}$ & $ 9.596\pm^{ 0.467}_{0.362}$\\
G34/W33  & 11:18:25.86  $+$58:46:42.6 & $ -9.53$ & $23.175\pm0.015$ & $  1.072\pm0.019$ & $1.427\pm0.074$ & $ 8.656\pm^{0.496}_{1.093}$ & $-0.639\pm^{0.094}_{0.049}$ & $10.175\pm^{ 0.330}_{0.683}$\\
G46/W40  & 11:18:23.69  $+$58:46:11.7 & $ -9.11$ & $23.598\pm0.019$ & $  1.106\pm0.025$ & $1.746\pm0.069$ & $ 4.504\pm^{1.414}_{1.161}$ & $-0.130\pm^{0.104}_{0.171}$ & $ 5.344\pm^{ 0.795}_{0.657}$\\
$\cdots$ & $\cdots$ & $\cdots$ & $\cdots$ & $\cdots$ & $\cdots$ & $\cdots$ & $\cdots$ & $\cdots$ \\
\hline
\multicolumn{9}{c}{\bf NGC\,3377}\\
G1   & 10:47:42.39 $+$13:59:33.7& $-10.03$ & $20.142\pm0.006$ & $1.136\pm0.008$ & $1.779\pm0.015$ & $ 7.039\pm^{4.454}_{2.178}$ & $-0.091\pm^{0.088}_{0.263}$ & $18.439\pm^{5.426}_{2.662}$\\
G2   & 10:47:43.74 $+$14:00:17.5& $ -9.87$ & $20.297\pm0.006$ & $0.964\pm0.008$ & $1.226\pm0.020$ & $13.703\pm^{5.873}_{9.113}$ & $-1.100\pm^{0.370}_{0.946}$ & $15.579\pm^{4.960}_{7.703}$\\
G4   & 10:47:39.23 $+$13:59:42.6& $ -9.63$ & $20.538\pm0.007$ & $0.962\pm 0.01$ & $1.304\pm0.016$ & $ 9.278\pm^{9.626}_{5.452}$ & $-0.880\pm^{0.174}_{0.841}$ & $ 9.484\pm^{6.514}_{3.689}$\\
G5   & 10:47:43.74 $+$13:59:46.1& $ -9.45$ & $20.722\pm0.008$ & $1.153\pm0.011$ & $1.683\pm0.011$ & $ 9.635\pm^{0.595}_{4.688}$ & $-0.348\pm^{0.264}_{0.049}$ & $12.092\pm^{0.557}_{4.348}$\\
G6   & 10:47:37.31 $+$14:00:04.5& $ -9.42$ & $20.753\pm0.007$ & $0.993\pm 0.01$ & $1.293\pm0.015$ & $ 9.910\pm^{9.195}_{6.043}$ & $-0.906\pm^{0.196}_{0.918}$ & $ 8.132\pm^{5.101}_{3.358}$\\
$\cdots$ & $\cdots$ & $\cdots$ & $\cdots$ & $\cdots$ & $\cdots$ & $\cdots$ & $\cdots$ & $\cdots$ \\
\hline
\multicolumn{9}{c}{\bf NGC\,584}\\
G1    & 01:31:19.98  $-$06:51:40.5 & $-10.63$ & $20.891\pm0.007$ & $0.623\pm0.011$ & $0.726\pm0.032$ & $ 0.005\pm^{0.001}_{0.001}$ & $-0.933\pm^{0.464}_{0.863}$ & $ 0.106\pm^{ 0.015}_{0.030}$\\
G2    & 01:31:20.46  $-$06:52:50.1 & $-10.31$ & $21.207\pm0.008$ & $1.097\pm0.011$ & $ 1.67\pm0.012$ & $ 5.906\pm^{0.091}_{0.761}$ & $-0.382\pm^{0.110}_{0.015}$ & $18.142\pm^{ 0.234}_{1.611}$\\
G4    & 01:31:17.07  $-$06:52:25.1 & $-10.16$ & $21.356\pm 0.01$ & $0.597\pm0.016$ & $0.828\pm0.044$ & $ 0.789\pm^{0.420}_{0.613}$ & $-0.804\pm^{0.344}_{0.765}$ & $ 2.207\pm^{ 1.133}_{1.646}$\\
G5    & 01:31:16.93  $-$06:51:21.1 & $ -9.61$ & $21.912\pm0.012$ & $ 1.35\pm0.016$ & $1.365\pm0.020$ & $20.000\pm^{0.000}_{0.000}$ & $-0.885\pm^{0.173}_{0.793}$ & $19.308\pm^{ 2.554}_{1.521}$\\
G6    & 01:31:17.28  $-$06:52:37.2 & $ -9.59$ & $21.935\pm0.013$ & $1.223\pm0.018$ & $1.984\pm0.020$ & $12.832\pm^{1.334}_{6.470}$ & $ 0.074\pm^{0.233}_{0.069}$ & $23.026\pm^{ 1.990}_{9.543}$\\
$\cdots$ & $\cdots$ & $\cdots$ & $\cdots$ & $\cdots$ & $\cdots$ & $\cdots$ & $\cdots$ & $\cdots$ \\
\hline\hline
\end{tabular}
\end{table}

\end{landscape}

\subsection{Completeness in K-band}\label{Sect:completeness}

\begin{figure}
\epsfig{file=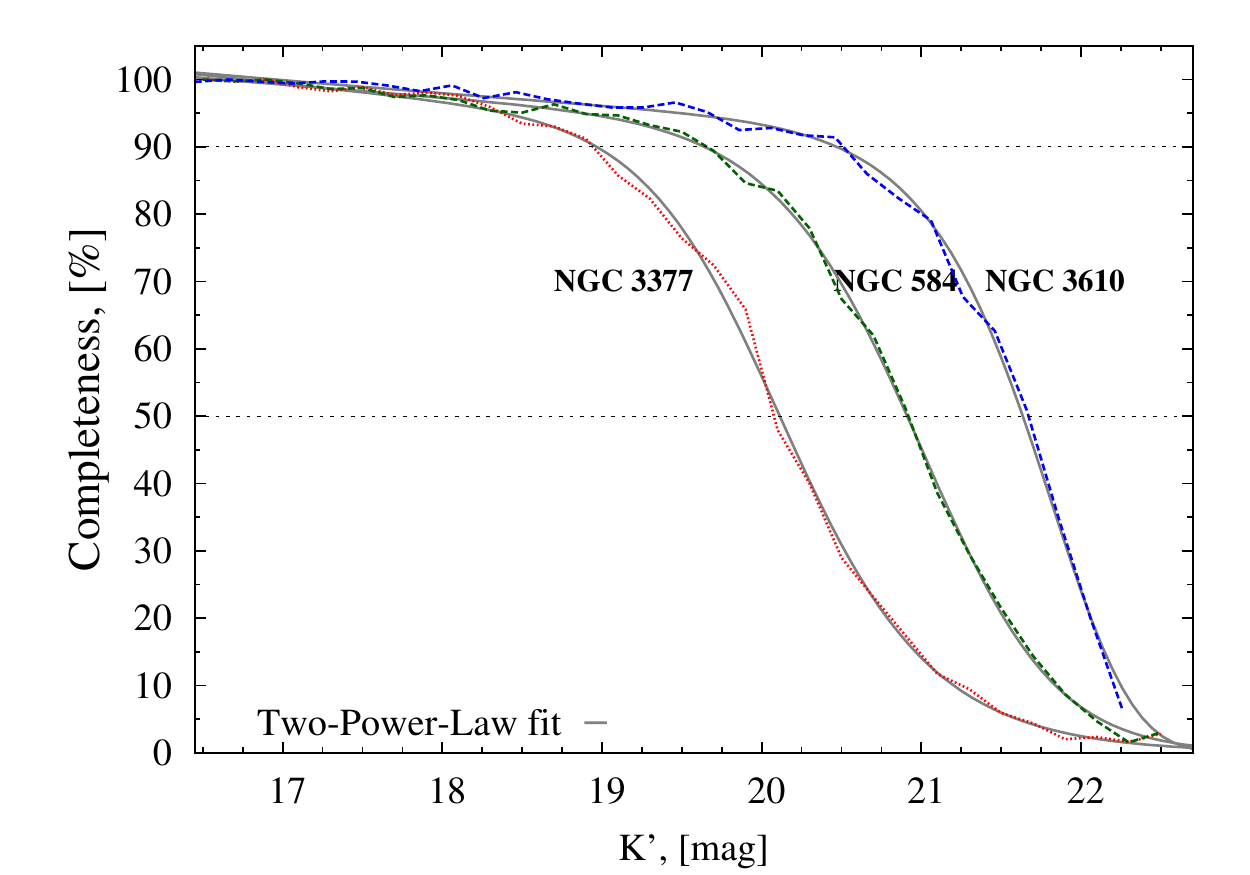, width=0.5\textwidth, bb=40 10 370 250}
\caption{Completeness curves for object detection as a function 
of $K^\prime$ magnitude for the three studied galaxies. Dotted 
horizontal lines mark the $50\%$ and $90\%$ completeness limits 
shown in Table\,\ref{table:completeness}. Smooth solid lines show 
the two-power-law fit to the completeness distributions.
}\label{fig:completeness}
\end{figure}

To determine the completeness of the K-band observations, 
we used the {\sc addstar} task in IRAF to add artificial sources 
to the final stacked image of each galaxy. 
To avoid introducing increased crowding to the images while retaining 
a high number of artificial stars per spatial and magnitude bin 
for reliable statistics, we added 100 stars per image and repeated 
this step 100 times\footnote{The value of the seed in {\sc addstar} 
was different for each frame}, i.e. in total $10^4$ artificial 
stars per galaxy. The images were then run through the same 
extended galaxy light subtraction, detection and photometry steps. 
Detected and input artificial star lists were matched to compute 
the completeness as a function of $K^\prime$ magnitude. The 
completeness curves are shown in Figure\,\ref{fig:completeness}. 
With smooth solid lines we show a two-power-law fit to the completeness 
values at each 0.5\,mag bin. These functions were later used to 
assess completeness levels as a function of age and metallicity 
(see Sect.\ \ref{Subsect:AgeZ Distributions})

In Table\,\ref{table:completeness} we summarize the $50\%$ and $90\%$ 
completeness values for the three galaxies. The absolute $V$-band 
magnitude ($M_V$) was calculated using the galaxies' respective distance moduli
(see Sect.\,\ref{SubSect:Observations}), and a $V-K^\prime=2.5$\,mag 
colour for the GCs, which is the mean of the observed range of 
$2.0\lesssim V-K^\prime\lesssim3.0$\,mag. Thus, at the respective 
distance to these galaxies we observe, with $50\%$ completeness, 
just at the GC luminosity function turnover magnitude (GCLF ToM) for 
NGC\,3377, and about 0.5 and 1\,mag brighter than the GCLF ToM for 
NGC\,584 and NGC\,3610, respectively. This corresponds to a luminosities greater 
than a few\,$10^5L_\odot$, meaning that our $K^\prime$-band GC 
detection samples the brightest and most luminous GCs.  We note, 
that the currently best estimate of the distance modulus of NGC\,3610 was  
derived $\sim3$\,yrs after the observations (see Sect.\ \ref{SubSect:Observations}), 
which were planned with a 
closer distance to NGC\,3610. 
This led to a lower completeness than planned for NGC\,3610. 
\begin{table}
\centering
\caption{Completeness limits for object detection from $K^\prime$-band 
completeness test. $M_V$ calculated using a mean $V-K^\prime=2.5$\,mag 
colour.
}\label{table:completeness}
\begin{tabular}{l|ccc}
\hline\hline
Target & $K^\prime$ & $M_V$ & $L_V$\\
& [mag] & [mag] & [$10^5L_{V,\odot}$]\\
\hline\hline
&\multicolumn{3}{c}{$50\%$ completeness}\\

NGC 3610 & 21.71 & $-8.50$ & $2.1$ \\
NGC\,584 & 20.92 & $-8.12$ & $1.5$ \\
NGC\,3377 & 20.08 & $-7.59$ & $0.92$\\
\hline
&\multicolumn{3}{c}{$90\%$ completeness}\\

NGC 3610 & 20.51 & $-9.70$ & $6.4$ \\
NGC\,584 & 19.66 & $-9.36$ & $4.7$ \\
NGC\,3377 & 18.95 & $-8.72$ & $2.6$\\
\hline\hline
\end{tabular}
\end{table}

\section{$VIK^\prime$ colour, age and metallicity distributions of GCs}\label{Sect:VIK colour-colour plot, ages and Zs} 

As already mentioned in the Introduction, the combination of 
optical and near-infrared imaging photometry 
is a very efficient method to access the distribution of ages and 
metallicities for an entire globular cluster system of a galaxy by means of 
a comparison to predictions of stellar population synthesis models. 
The specific power of this "optical+NIR" method is twofold: 
\begin{enumerate}
\item for a population of coeval stars older than $\gtrsim1$\,Gyr,
 near-IR colour indices like $I\!-\!K$ or $J-K$ mainly sample the metallicity   
 sensitive temperature distribution of stars on the RGB \cite[e.g.][]{Worthey94b}. 
 In combination with an optical colour like $V\!-\!I$, which is sensitive to both
 metallicity and age, this allows to break down the age-metallicity degeneracy 
 present in the  optical colour indices. 
\item Relative to optical colours and spectral indices, near-IR colours are
 significantly less sensitive to the presence of {\it evolved\/} hot stars
 such as blue horizontal branch (HB) stars and blue stragglers. Blue HBs can
 be present in GCs older than $\sim8$\,Gyr \citep[starting with metal-poor GCs; see
 e.g.][]{LeeYoonLee00,Yi03}, and the relative number of blue
 versus red HB stars in GCs is known to depend mainly on age,  
 metallicity, and GC mass \citep[e.g.,][]{Carretta10,Dotter10,Dotter11}. 
Hence, the presence of blue HB stars in GCs can
 cause incorrect age determinations when using optical photometry or
 spectroscopy \citep[see also][]{Schiavon04,Cenarro07,Percival&Salaris11}, 
while NIR colours are largely insensitive to their presence. 
\end{enumerate}
In the following, our age-metallicity analysis will be based on $V\!-\!I$
versus $I\!-\!K'$ diagrams. 

\begin{figure*}
\epsfig{file=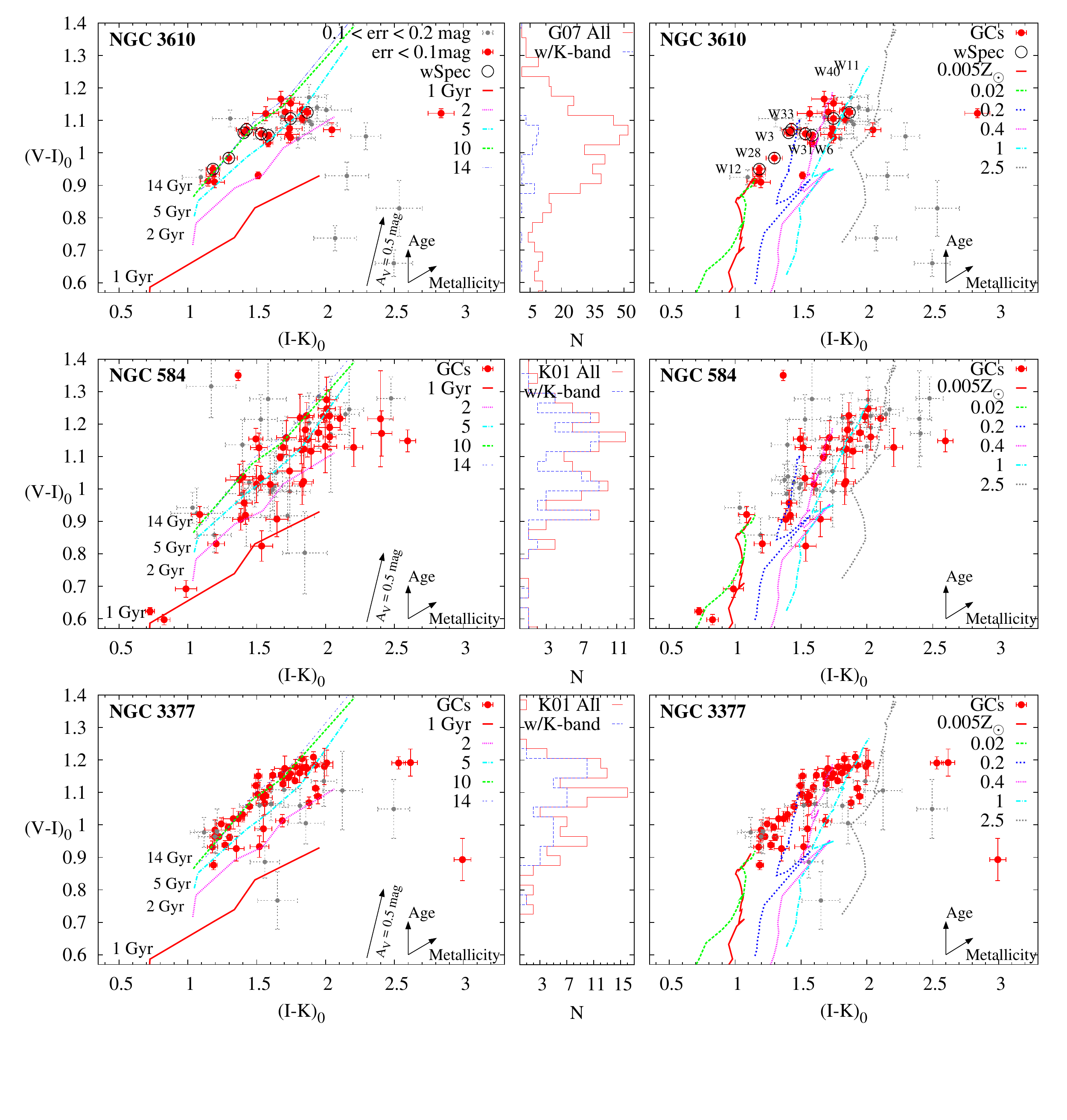, width=1\textwidth, bb=20 70 700 756}
\caption{Optical-NIR $(V\!-\!I)_0$ vs. $(I\!-\!K)_0$ colour-colour distributions 
of globular clusters (solid circles) in NGC\,3610, 584 and 3377 from top 
to bottom, respectively. Encircled solid symbols mark GCs with spectroscopic 
measurements in NGC\,3610 \citep{Strader03,Strader04}. Small dots are GCs with colour 
error greater than 0.1\,mag. colour indices are foreground reddening corrected. 
{\bf Middle panels} show histograms of $(V\!-\!I)_0$ colour distributions of the 
GC population detected with HST (with solid line) and those with NIRI 
$K^\prime-$band magnitudes (dashed line). Smaller solid circles with 
dashed line error bars show GCs with colour uncertainty larger than 0.1\,mag. 
{\bf Left} and {\bf right} panels show a comparison between cluster 
colours and expectations for ages and metallicities from Simple Stellar 
Population (SSP) synthesis models \citep{BC03}. Different line-types 
indicate isochrones of different ages for all metallicities covered by 
the models (left panels) and isometallicity SSP tracks for ages $>0.5$\,Gyr 
(right panels). The arrows in the bottom panels show the direction of 
increasing age and metallicity, as well as a reddening vector 
for $A_V=0.5$\,mag.
}\label{fig:VI-IK}
\end{figure*}
In Figure\,\ref{fig:VI-IK}, we present the $V\!-\!I$ vs. $I\!-\!K^\prime$ 
colour-colour distributions of all GCs detected in $K'$ in 
our sample galaxies. Those are compared to \cite{BC03} SSP models 
using a canonical \citet{Chabrier03} IMF, 
for several ages (1, 2, 5, and 14 Gyr) and 
metallicity values ($Z/Z_{\odot}$ = 0.005, 0.02, 0.2, 0.4, 1.0, and 2.5,
equivalent to [Fe/H] = $-$2.3, $-$1.7, $-$0.7, $-$0.4, 0.0, and
+0.4)\footnote{We note that the quoted photometric [Fe/H] is in fact [Z/H],
  i.e. the total content of metals relative to hydrogen given in the SSP
  models. However, the difference between [Z/H] and [Fe/H] is 0.03\,dex
  $<\Delta$[Fe/H]$<0.06$\,dex for low to high metallicities, i.e. much
  smaller than the measurement error. To avoid confusion in the following we
  adopt the [Fe/H] annotation for the photometric metallicities as well.}. 
Figure~\ref{fig:VI-IK} shows that the three galaxies generally show a large  
spread in age and metallicity among their brightest GCs. A notable 
subpopulation of roughly coeval bright GCs can be also seen for NGC\,3377 
(see below). 
The $I\!-\!K'$ colours of the majority of the "blue" population of GCs as judged
from the optical data (i.e., $V-I<1.0$) 
%
turn out to be consistent with old ages ($\gtrsim5$\,Gyr) and
low-metallicities on average ($-2.3\lesssim[$Fe/H$]\lesssim-0.7$\,dex), while
the $I\!-\!K'$ colours of the optically redder clusters tend to indicate higher
metallicities and younger ages ($[$Fe/H$]\gtrsim-0.7$\,dex and $\sim 3 -
6$\,Gyr, respectively). 

While the overall colour-colour distributions are indicative of a 
large spread in age and metallicity among the brightest clusters, 
there is a discernible population of roughly coeval GCs in NGC\,3377 (i.e., the 
GCs between the 2 and 5\,Gyr isochrones in the lower left panel of
Fig.\,\ref{fig:VI-IK}). From our data alone, there is no indication 
for a spatial correlation among these clusters (cf Fig.\ref{fig:N3610_N584_N3377_mos2}). 
While such a subpopulation is not readily seen in the other 
two galaxies, this may be due at least in part to the photometric
completeness limits (cf. Table\,\ref{table:completeness}). The middle panel
histograms in Figure\,\ref{fig:VI-IK} show that our NIRI imaging detects 
$61\%$ (64/106), $73\%$ (81/115) and $8\%$ (50/611) of the 
NGC\,3377, NGC\,584 and NGC\,3610 GCs detected with HST, respectively. 
Although the NIRI imaging samples $73\%$ of the NGC\,584 GCs observed 
with WFPC2, the relatively large photometric errors cause a larger scatter 
in their $VIK'$ colour distributions, which prohibits a clear 
detection of any underlying subpopulation of coeval GCs similar 
to that observed in NGC\,3377. 

\subsection{Photometric ages and metallicities}\label{Subsect:Photometric Age Z}

Employing the good resolution in age and metallicity provided by the 
optical-NIR ($V,I,K$) colour indices, we derive photometric ages and 
metallicities for the GCs in our sample. This is done by a standard 
$\chi^2$ minimization interpolation between SSP model tracks of 
\citet{BC03}. A brief comparison with other SSP models 
and the reasoning why we chose \citeauthor{BC03} models for this purpose is
presented in Section\ \ref{Sect:SSPs  comparison}. The derived ages and
metallicities are the weighted  
average of the age and metallicity of the two nearest tracks. The 
weights include the distance to the respective model and the photometric 
errors. This way we also calculate an upper and lower error to the 
photometric ages and metallicities. We list the derived properties of all GCs
in our sample in Table\,\ref{Table:VIK_ageZM}. 

\begin{figure}
\epsfig{file=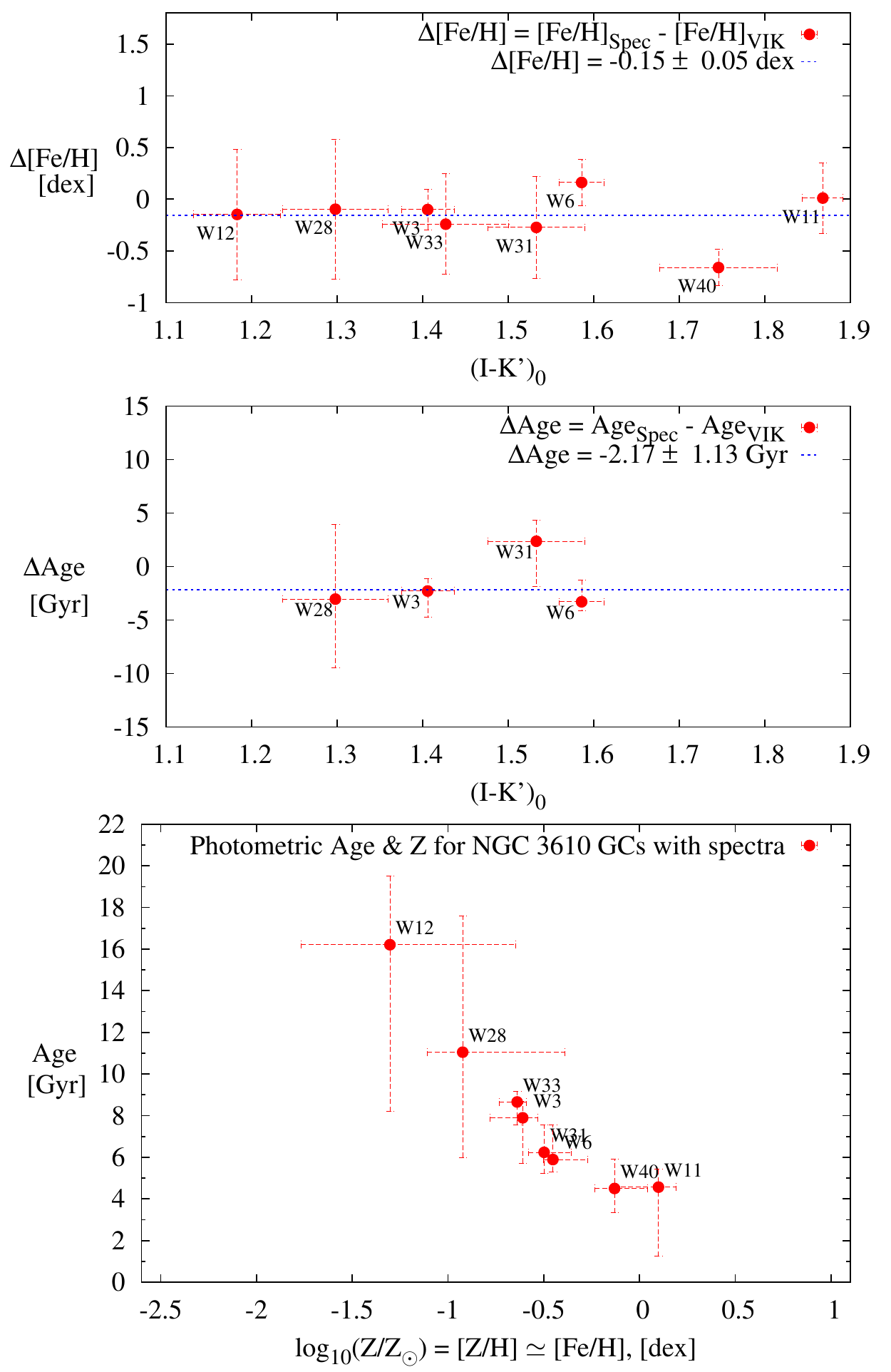, width=.5\textwidth, bb=20 10 360 552}
\caption{Differences between spectroscopic and photometric metallicities 
({\bf top}) and ages ({\bf middle}) for NGC\,3610 GCs. Dashed lines are 
the least squares fit. {\bf Bottom} panel: photometric ages and metallicities 
for NGC\,3610 GCs with spectra. The top panel shows that VIK derived 
metallicities are well consistent with their spectroscopic values. On 
average, photometric ages are older than spectroscopic ones by $\sim2$\,Gyr 
(middle panel), and the difference between them is more pronounced for 
low-metallicity GCs than those with higher $Z$s, possibly due to the 
presence of blue horizontal branch (see text in Sect.\,\ref{Subsect:Photometric Age Z}).
}\label{fig:ageZ_difference}
\end{figure}

About a dozen of the brightest GCs in NGC\,3610 have 
available Lick index analysis of optical spectroscopy 
\citep{Strader03,Strader04}. \citeauthor{Strader03} qualitatively 
discuss the clusters ages, but they do not explicitly provide ages 
for each cluster. Thus, using their Lick index strengths, we calculate ages 
for the GCs with the {\sc gonzo} code \citep{Puzia02b}, using the models of
\citet{Thomas03}. Of those GCs, eight are within the NIRI field of view for
which we can compare our ages and metallicities with their spectroscopic values. 
In Figure\,\ref{fig:ageZ_difference}
we show these differences (in the sense spectroscopic minus photometric) as a
function of $I\!-\!K'$ colour. Errors in the top and middle panels in
Figure \ref{fig:ageZ_difference} are the square root of the sum of the
squares of the spectroscopic and photometric errors while errors in the bottom
panel are their photometric uncertainties. 

The least squares fit to the data (dashed lines in Fig.\,\ref{fig:ageZ_difference}) 
show that spectroscopic [Fe/H] and photometric metallicities [Z/H] 
derived from the $VIK$ colours agree very well, keeping in mind that 
[Fe/H$]-[Z/$H]$=0.03$ to 0.06\,dex 
(cf.\ Sect.\,\ref{Sect:VIK colour-colour plot, ages and Zs} above). 
In contrast, the photometric age estimates of the most massive GCs in NGC\,3610 are 
older than their spectroscopic values by about 2 Gyr on average. The age
difference is larger for low-metallicity GCs than for higher metallicity ones. Such 
a difference can in principle be caused by the presence of blue horizontal
branch (BHB), helium-core burning stars in these clusters. It is well known 
that spectroscopic luminosity-weighted ages, derived from the classical 
H$_\beta$ vs Mg2 or Mgb absorption line indices, can be biased toward 
younger ages, especially for metal-poor GCs, because the high effective 
temperature of BHB stars ($T_{eff}>9000$\,K) and blue stragglers enhances 
the strength of the Balmer lines
\cite[e.g.][]{LeeYoonLee00,Schiavon04,Koleva08,Ocvirk10,Percival&Salaris11,Xin11}.  
An example for such "age bias" are the metal-poor Galactic GCs NGC\,3201
and NGC\,5024 featuring BHBs, whose spectroscopically derived age (from
H$_\beta$ vs.\  Mg$_2$) is $\sim8$\,Gyr \citep{Perina11}, while their ages
from deep HST CMD fitting is $\gtrsim12$\,Gyr \citep{Dotter10}. 
Furthermore, nearly all 
massive (${\cal M}\gtrsim3\times10^5M_\odot,\ M_V<-8.9$\,mag) Galactic 
GC have hot HBs \cite[e.g.][]{Recio-Blanco06,Lee07}, with the exception 
of the metal-rich GC 47\,Tuc. This may be due to its low central 
escape velocity ($\upsilon_{\rm esc}$) for a massive cluster at that metallicity, 
unlike the metal-rich NGC\,6388 and NGC\,6441 which do have BHBs and 
larger $v_{\rm esc}$ \cite[cf.\,\protect\href{http://tinyurl.com/g09fig5}{Fig.\,5}
in][]{Georgiev09b} and thus better able to retain/accrete processed stellar
ejecta. \cite{Puzia02b} performed a Lick index analysis for metal-rich
Galactic GCs and they note that NGC\,6388 and NGC\,6441 (which host BHB
stars) show a stronger H$_\beta$ index than the other metal-rich RHB GCs. 
However, \cite{Schiavon04} estimate ages of $\sim8$ and 
$\sim11$\,Gyr from Lick index spectroscopy for these two clusters, which
indicates that spectroscopic age-dating of metal-rich GCs can be less biased 
by the HB morphology. This is what we do indeed observe in 
the middle panel of Figure\,\ref{fig:ageZ_difference} for W\,6 and 
W\,31, two metal-rich GCs in NGC\,3610: their photometric and spectroscopic  
ages are more consistent (see also Sect.\,\ref{Sect:CMD-masses}). More pronounced HB effect is expected for 
metal-poor GCs, which we indeed observe for the two metal-poor GCs 
in the middle panel of Figure\,\ref{fig:ageZ_difference}.
This indicates that in spite of uncertainties inherent to SSP models, the most
likely cause for the mean difference between photometric and spectroscopic
ages for these GCs is the presence of hot HB stars, in \emph{particular} for
the metal-poor GCs. 

\begin{figure}
\epsfig{file=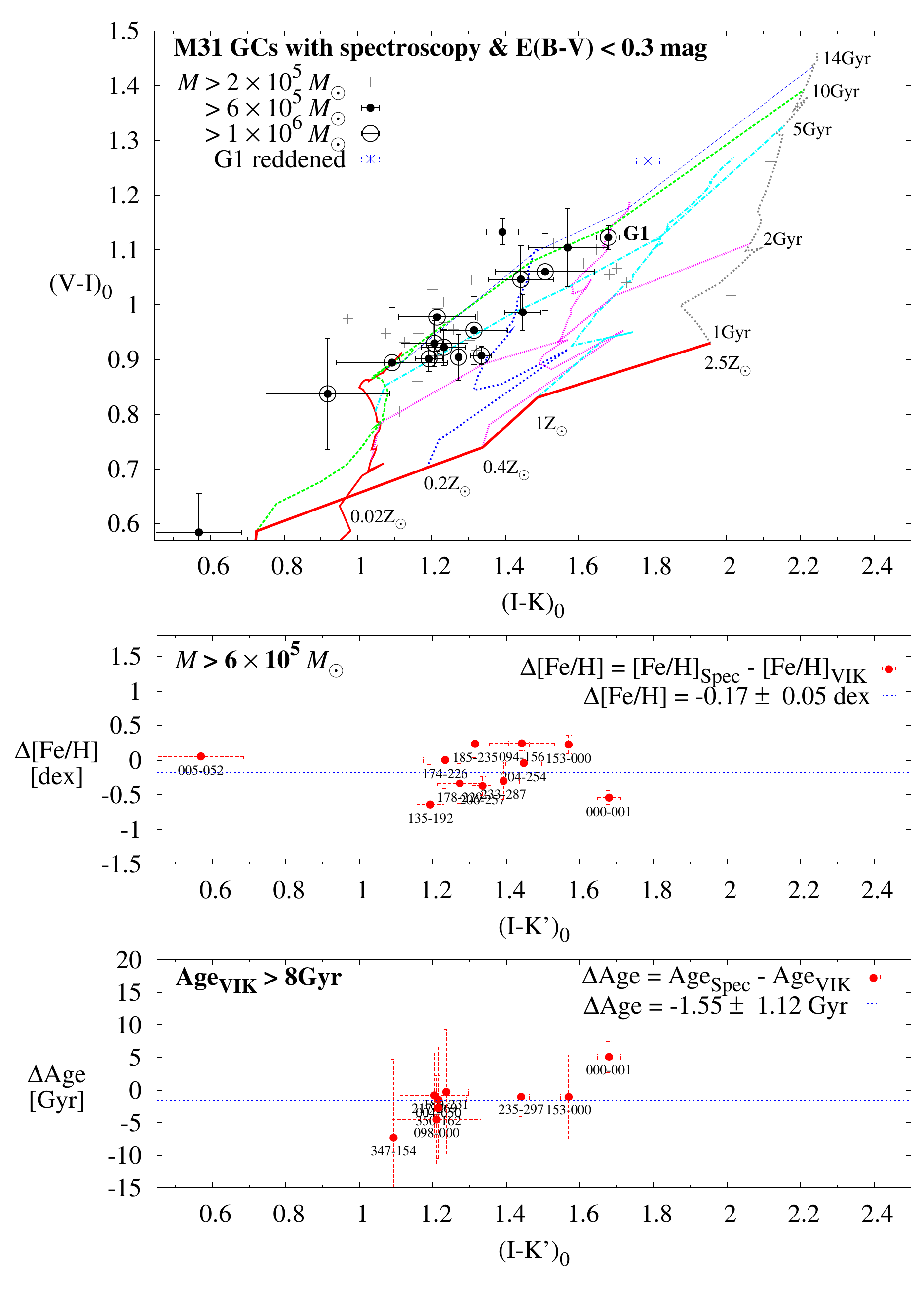, width=.35\textwidth, bb=20 30 400 750}
\caption{Comparison between $VIK$ photometric and spectroscopic ages and 
metallicities for M\,31 GCs. {\bf Top}: Brightest (most massive) and less 
reddened ($E(B\!-\!V)\!<\!0.3$\,mag) M\,31 GCs, as indicated in the legend, 
which we consider having the most reliable photometric and spectroscopic measurements. 
$VIK$ magnitudes are from \protect\cite{Barmby00,Barmby01} and spectroscopic ages and 
metallicities from \protect\cite{Beasley05} and \protect\cite{Caldwell11}. Colours were 
dereddened using $E(B\!-\!V)$ values from these studies. In spite of the large 
photometric and reddening uncertainties, we derive similar photometric and 
spectroscopic metallicities ({\bf middle panel}), while spectroscopic 
ages for the oldest GCs (Age$_{VIK}>8$\,Gyr) are younger on average ({\bf bottom panel}), 
signaling the BHB effect. In both panels, dashed horizontal lines are the 
least-squares fits to the data, with coefficients shown in the legend.
}\label{fig:M31 GCs VIK AgeZ}
\end{figure}

The validity of the method of deriving ages and metallicities from $V,I$, 
and $K$ photometry used in this paper is supported by the observation that 
dereddened $V\!-\!I$ and $V\!-\!K$ colours of old GCs in the Milky Way and M\,31
have been shown to be consistent with \citeauthor{BC03} SSP model predictions 
for "old" ages ($\ga 10$\,Gyr) throughout the range of [Z/H] sampled, 
\cite[see Fig. 5 of][]{Puzia02}. The applicability of our method is illustrated 
in Figure\,\ref{fig:M31 GCs VIK AgeZ} using the same M31 GCs data 
as in \cite{Puzia02}, i.e. $V,I$ magnitudes from \cite{Barmby00,Barmby01}, 
augmented with new spectroscopic measurements from \cite{Beasley05} and 
\cite{Caldwell11}, the latter providing updated $E(B\!-\!V)$ values. Even though 
the $E(B\!-\!V)$ are uncertain \cite[$\sigma\!\sim\!0.1$\,mag, see][]{Caldwell11}, 
the $V\!-\!I,I\!-\!K$ colours are consistent with the \citeauthor{BC03} SSP model 
predictions, even so for GCs massive ($M>6\times10^5 M_\odot$) and 
$E(B\!-\!V)\!<\!0.3$\,mag. We derive their photometric ages and metallicities, which 
compare well with their spectroscopic values from \cite{Caldwell11}, while 
spectroscopic ages for the oldest GCs (Age$_{VIK}\!>\!8$\,Gyr), on average, are younger, 
i.e. showing the BHB effect. Due to this problem, \cite{Caldwell11} does not 
derive spectroscopic ages for the majority of the GCs and assume age of 14\,Gyr. 
We consider only GCs with spectroscopic ages in Figure\,\ref{fig:M31 GCs VIK AgeZ}.

\subsection{Comparison to various SSP models}\label{Sect:SSPs comparison}

\begin{figure*}
\epsfig{file=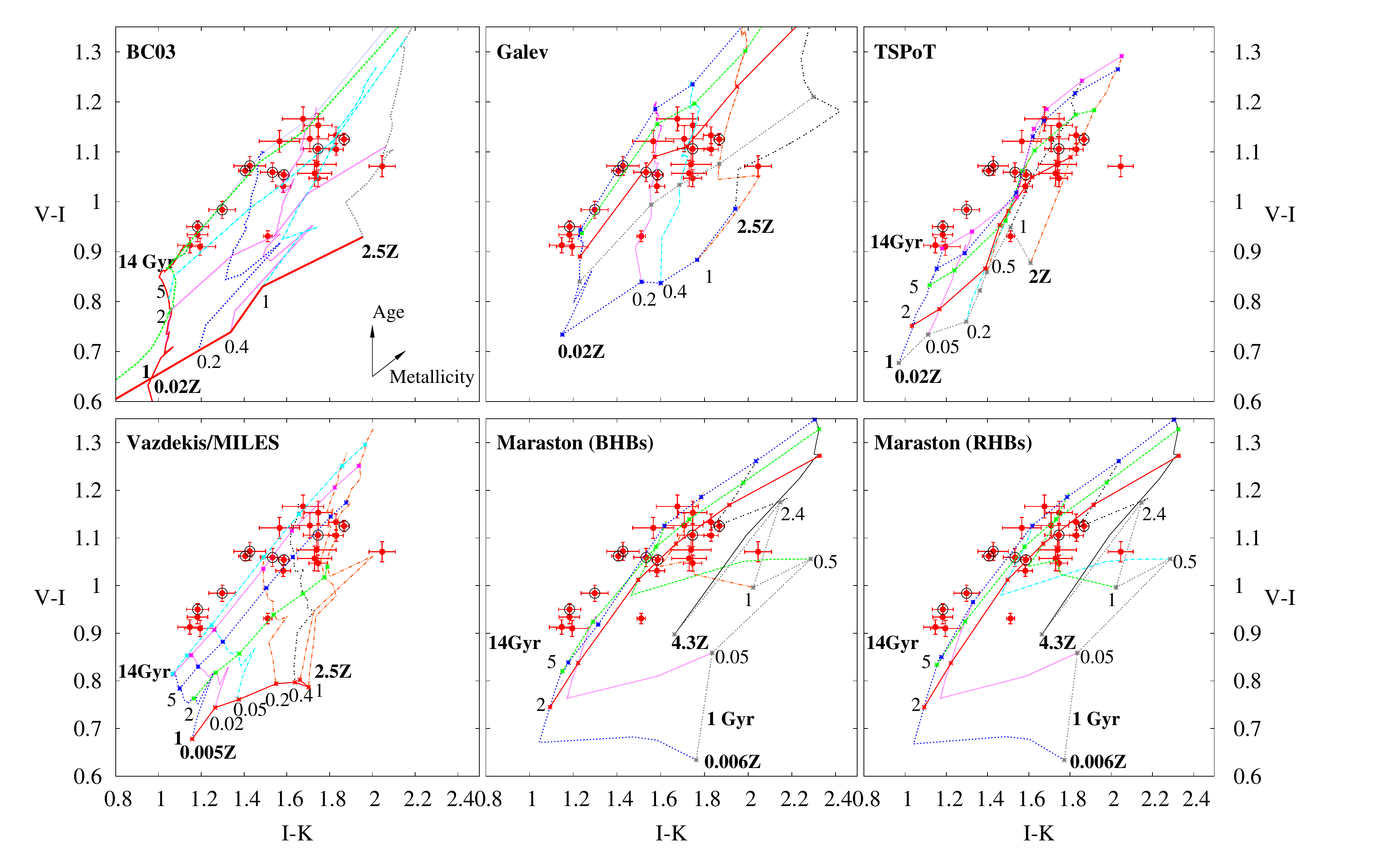, width=1.1\textwidth, bb= 20 20 1080 650}
\caption{Comparison between $V-I,I-K^\prime$ colours of GCs in NGC\,3610 
with colour errors $<0.1$\,mag and different SSP synthesis models (see 
text in Sect.\,\ref{Sect:SSPs comparison}). With circles are GCs which have 
spectroscopically derived ages and metallicities. In the 
top left panel with arrows are indicated the direction of increasing age 
and metallicity. With labels are indicated the $1,2,5$ and $14$\,Gyr 
isochrones and isometallicity SSP tracks with labeled values. All models 
are with canonical IMF \citep{Kroupa01,Chabrier03}. MILES SSPs are with 
the revised Kroupa IMF, which accounts for effects from unresolved 
binaries. Only \citeauthor{BC03} SSP models provide $K^\prime$ magnitudes, 
however, the $K-K^\prime$ difference is from $-0.02$ to 0.007\,mag for 
high to low metallicities, respectively, which is negligible to account 
for differences between data and models. The Maraston models include the 
blue and intermediate-red horizontal branches tracks, which differ for 
ages $>5$\,Gyrs. 
}\label{fig:SSPs}
\end{figure*}

It is beyond the aims of this paper to test different SSP models and 
investigate the nature of the differences between them. However, in 
order to provide a qualitative overview of how different synthesis 
population models compare to the observed colours of NGC\,3610 GCs with available
spectroscopy, we compare our data with the popular SSP models of BC03
\citep{BC03}, Galev \citep{Anders03,Anders09}, Vazdekis/MILES
\citep{Vazdekis10}, TSPoT \citep{Brocato00,Raimondo05}, 
and the Maraston \citep{Maraston05} SSPs for blue and intermediate-to-red HBs. Results
are shown in Figure\,\ref{fig:SSPs}. 
As seen, the different SSP models yield significantly different 
results in terms of ages and metallicities. This is due in part to the
different treatment of key stellar evolutionary phases, e.g. the TP-AGB and
post-AGB phases, 
HB stars, etc., which are relatively poorly understood and empirically
constrained. 
In addition, the different SSP models use different 
stellar atmosphere spectral libraries (theoretical or empirical), and
different stellar evolution prescriptions. The effects of different choices of
the IMF on the integrated $VIK'$ colours are, however, smaller than the
observational errors. 
Overall, our $VIK'$ data is best represented by the BC03 SSP model, especially
for low metallicities and old ages. This is why we use BC03 
to derive and discuss ages and metallicities for our clusters in the
following. 

\subsection{GC age and metallicity distributions}\label{Subsect:AgeZ Distributions}

\begin{figure}
\epsfig{file=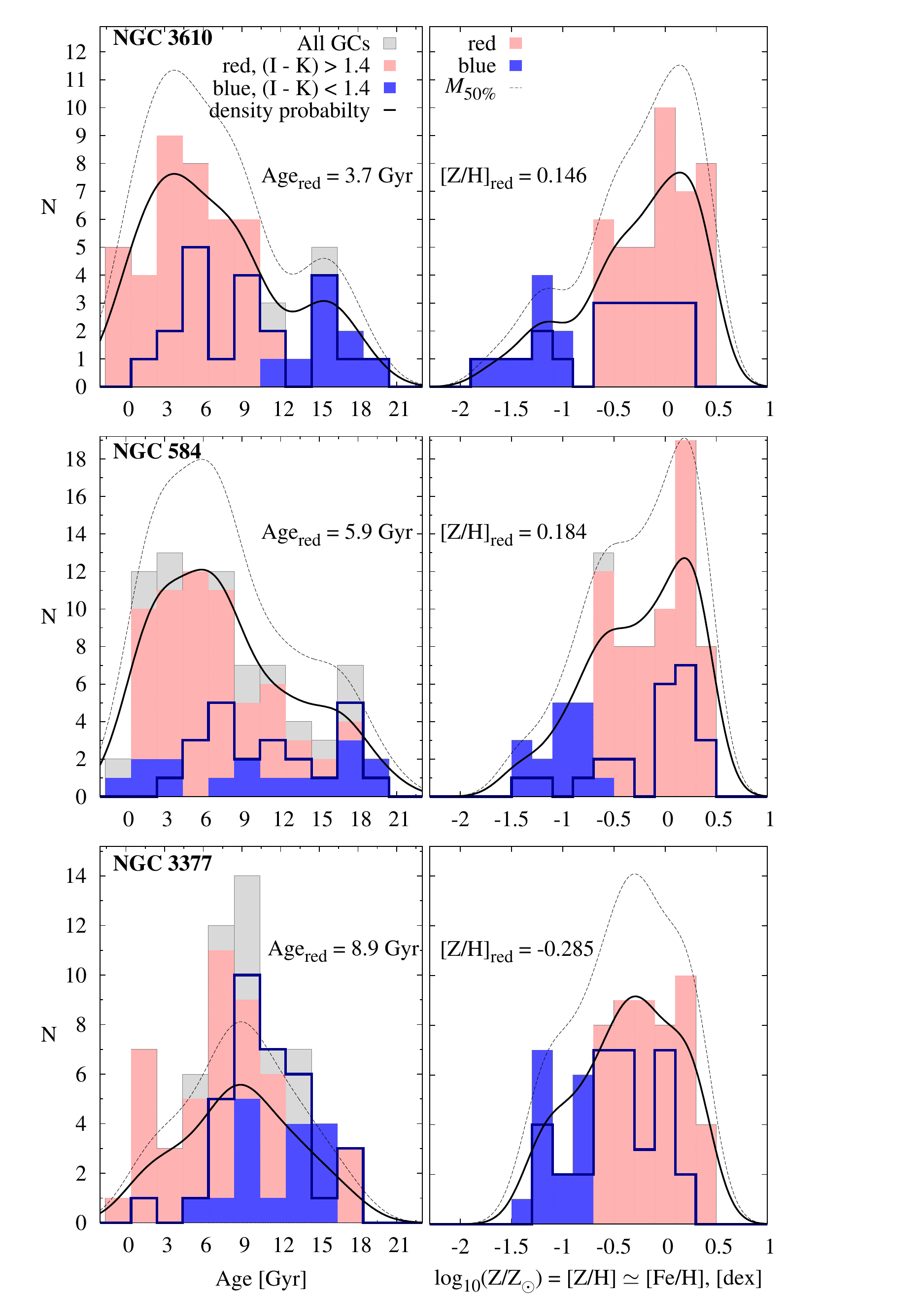, width=.35\textwidth, bb=50 10 340 750}
\caption{$VIK$ photometric age and metallicity distributions (left and 
right panels) for all GCs in the three post-merger remnant galaxies, 
i.e. NGC\,3610, NGC\,584 and NGC\,3377, from top to bottom respectively. 
Light solid (gray), solid (red) and dark solid (blue) histograms show 
all GCs, redder and bluer than $I\!-\!K^\prime=1.4$\,mag, respectively. 
Solid lines show the probability density distribution for all GCs. The 
highest probability density age for the red, metal-rich GCs, shown with 
label for each galaxy, increases from top to bottom. Dashed lines indicate 
the number density distribution corrected for $50\%$ incompleteness (details 
in text in Sect.\,\ref{Subsect:AgeZ Distributions}). Open solid line 
histograms present GCs with mass larger than the mass at $50\%$ incompleteness 
at 14\,Gyrs (cf. also Fig.\,\ref{fig:MassZ Distributions}).
}\label{fig:AgeZ Distributions}
\end{figure}

In Figure\,\ref{fig:AgeZ Distributions} we show the $VIK'$ photometric 
age and metallicity distributions for the GCs in the three target 
galaxies. 
Employing the good separation in the $I\!-\!K$ colour in the SSP models 
(cf e.g. Fig.\,\ref{fig:VI-IK}) between metal-rich and metal-poor and 
older versus younger populations, we separate the GCs in two subpopulations, 
namely redder and bluer than $I\!-\!K=1.4$\,mag, corresponding to 
$[Z/H]\sim-1$\,dex for a population older than about 5\,Gyr.
Within $R$\footnote{R is a language and environment for statistical computing
  and graphics (\protect\href{http://www.r-project.org}{http://www.r-project.org}).},  
we then estimate the highest-probability age for the metal-rich GCs using 
a non-parametric probability density estimator with a Gaussian kernel with 
size of 2\,Gyr. 
Those probability density 
estimates are shown with solid lines in Figure\,\ref{fig:AgeZ Distributions} 
and the peak values for the age distribution of the metal-rich GCs (Age$_{\rm red}$) 
are shown with labels. It is seen that Age$_{\rm red}$ increases from 
$\sim4$\,Gyr for NGC\,3610, to $\sim6$\,Gyr for NGC\,584 and $\sim9$\,Gyr 
for NGC\,3377. 
To assess the statistical level of similarity between the age distributions of
the three galaxies, we consider that one can not expect Gaussian (or any other 
anticipated functional form) for the cluster age distributions. Formally, this 
precludes the use of the Kolmogorov-Smirnov test which is commonly used for
comparing statistical similarity between two samples. We therefore 
use the Mann--Whitney U test\footnote{
The Mann--Whitney U test (a.k.a. Mann-Whitney-Wilcoxon rank-sum test or u-test) is a
{\it non-parametric} statistical hypothesis test which calculates a measure of
the difference between two {\it independent} samples of observations. It 
is similar to the t-test, but is more robust because it compares the 
sums of ranks and thus is less sensitive to the presence of outliers.}, which
yields that the distribution of GC ages in NGC\,3610 is different  
from those of NGC\,584 and NGC\,3377 with a probability of 75\% and 98\%, 
respectively. The age distribution of GCs in NGC\,584 differs from that in 
NGC\,3377 with 98\% probability. 
Thus, the brightest GCs in NGC\,3610, NGC\,584 and  NGC\,3377 have 
statistically significant different age distributions from one another 
and the peak age of the metal-rich subpopulation increases from NGC\,3610 to 
NGC\,584 and NGC\,3377, respectively. 

The age and metallicity distributions are affected by observational 
incompleteness. To assess the correction that likely needs to be applied, 
we used the functional relations between $K^\prime$, mass, $Z$ and Age 
using the \cite{BC03} SSP model. Since the HST/ACS and WFPC2 $V,I$-band 
photometry is $100\%$ complete for all $K^\prime$ detections, no additional 
$V,I$ completeness corrections are necessary. We fit 
power-law and logarithmic functions to the $K^\prime$ vs. [Fe/H] and age 
relations for mass at the $50\%$ completeness level at 2 and 14\,Gyr and for 
all metallicities, respectively, obtained from the \cite{BC03} SSP models 
(cf. lines in Fig.\ \ref{fig:MassZ Distributions}). The age 
and [Fe/H] distributions were then convolved with these functional relations 
to illustrate the completeness correction for $50\%$ incompleteness. These 
estimates are shown with light dashed lines in Figure\,\ref{fig:AgeZ 
Distributions}. Due to the relatively shallow slope of the $M-Z$ relation 
(cf. Fig.\,\ref{fig:MassZ Distributions}), the incompleteness affects equally 
metal-poor and metal-rich GCs, unless if metal-poor GCs tend to be larger 
and less dense which could bias their detection.

Keeping in mind photometric uncertainties, the age and metallicity distributions 
in Figure\,\ref{fig:AgeZ Distributions} seem to indicate the presence of 
multiple GC populations in the sample galaxies.

\section{Colour-Magnitude and mass-metallicity distributions}\label{Sect:CMD-masses}

\begin{figure}
\epsfig{file=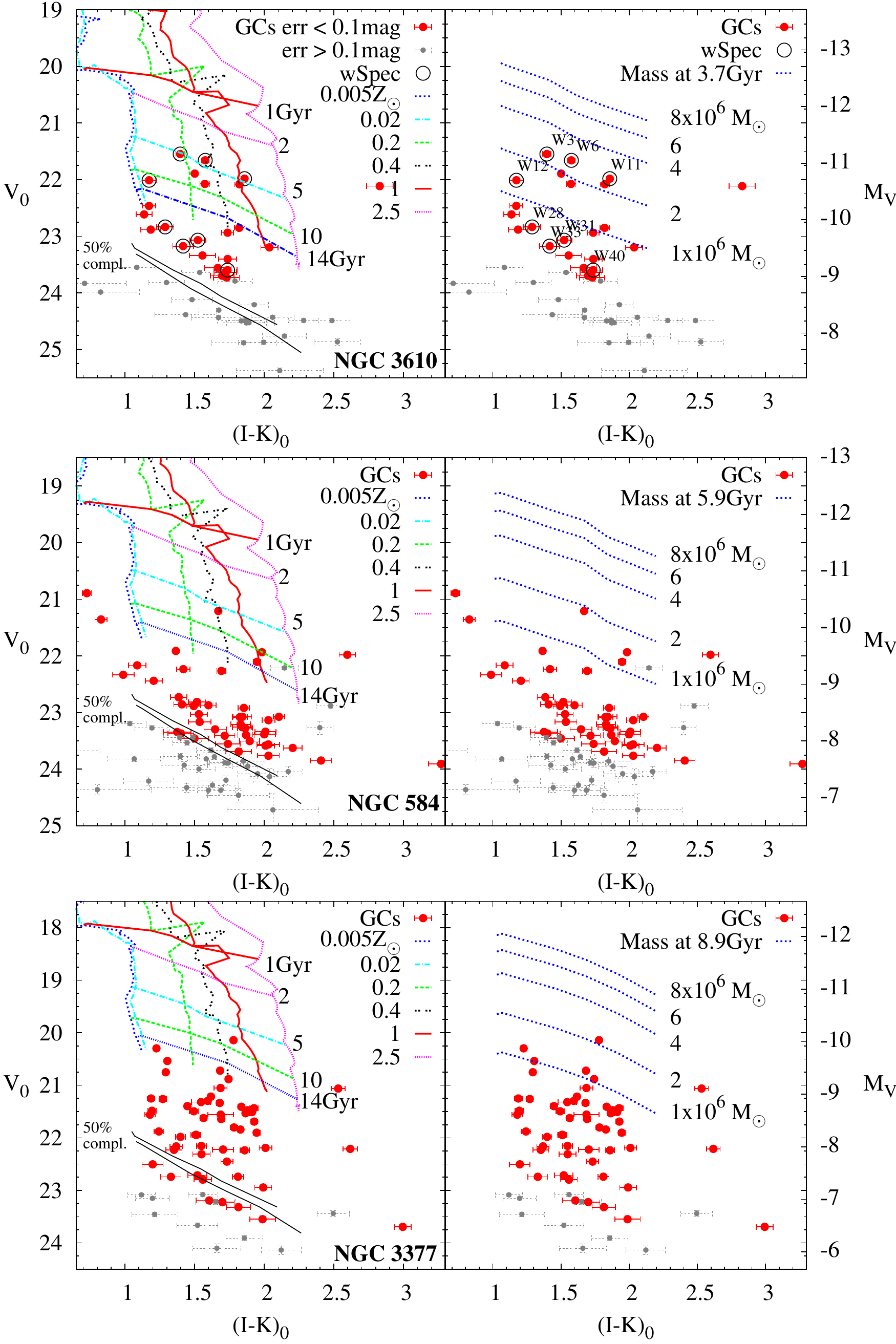, width=.6\textwidth, bb=20 10 600 750}
\caption[]{Colour-magnitude distribution of GCs (solid circles) in 
NGC\,3610, 584 and 3377, from top to bottom, respectively. Large 
circles are GCs with colour error $<0.1$\,mag. With circles in the 
NGC\,3610 panels are indicated GCs with spectroscopic observations 
in the literature \citep{Strader03,Strader04}. Lines in the {\bf 
left panels} show $1,2,5,10$ and 14\,Gyr isochrones and isometallicity 
\cite{BC03} tracks for $10^6M_\odot$ and $Z$ indicated in the legend. Lines in the 
{\bf right panels} mark a constant mass for all metallicities at 
GCs' peak age obtained from the probability density 
distributions in Fig.\,\ref{fig:AgeZ Distributions}, using 
the $M/L_V$-ratios provided in the models. 
The upper and lower solid lines in the left panels labeled with 
$50\%$ completeness indicate the $50\%$ completeness at 14 and 2\,Gyr.
}\label{fig:V-IK}
\end{figure}
The colour-magnitude diagrams of GCs in the three merger-remnant 
galaxies is presented in Figure\ \ref{fig:V-IK}. With different 
lines in the left-hand panels of Figure\ \ref{fig:V-IK} we show \cite{BC03} 
SSP model tracks for all metallicities scaled to the luminosity 
of a GC with ${\cal{M}} = 10^6$ M$_\odot$ at 14\,Gyr, using model
$M/L_V$ values. The iso-mass lines in the right-hand panels in
Figure\ \ref{fig:V-IK}  are calculated using the model $M/L_V$ at
14\,Gyr. This comparison shows that the masses of the GCs
with $K^\prime$ data range from a few $10^5M_\odot$
up to a few $10^6M_\odot$. 
The upper and lower solid lines in the left panels labeled with 
$50\%$ completeness indicate the $50\%$ completeness at 14 and 2\,Gyr,
respectively, estimated using the SSP model prediction for the $I\!-\!K$ colour.

\begin{figure}
\epsfig{file=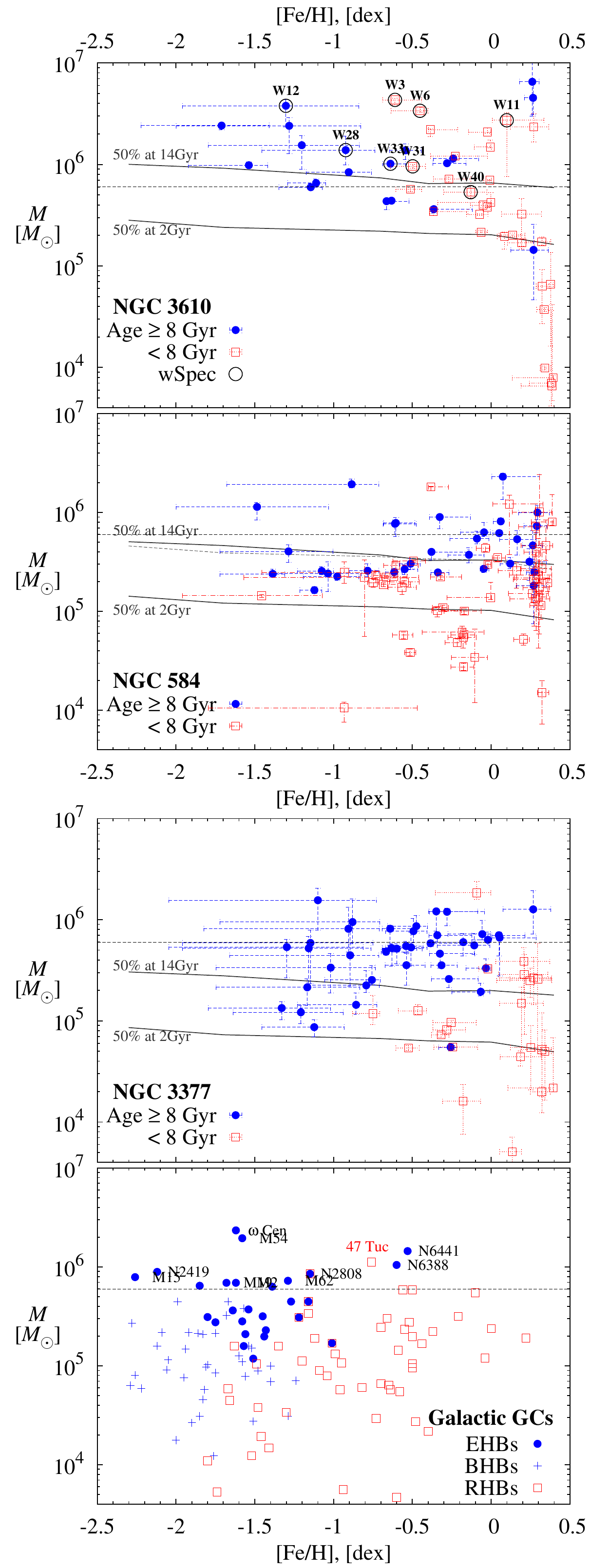, width=0.5\textwidth, bb=0 15 390 920}
\caption{Mass-metallicity distributions for GCs in NGC\,3610, 
NGC\,584, NGC\,3377 and in the MW (from top to bottom). Solid circles and 
open squares in the sample galaxies panels show GCs older and 
younger than 8\,Gyr, respectively. Solid symbols, crosses and 
open squares in the Galactic GCs panel separates GCs with very 
extended, blue and red horizontal branches. Dashed horizontal 
line marks the $6\times10^5M_\odot$, above 
which all Galactic GCs except 47\,Tuc have BHBs. Labeled solid lines indicate 
the $50\%$ completeness level at 2 and 14\,Gyr.
}\label{fig:MassZ Distributions}
\end{figure}

To convert the colour-magnitude diagrams to mass-metallicity distributions, 
we use $M/L_V$ values provided by the SSP model to convert from GC
luminosities to masses using the distance moduli to the galaxies. 
For comparison purposes, we also include info on GCs in the Milky Way
(hereafter MW GCs). 
MW GC masses were calculated using data from \cite{McLaughlin&vdMarel05} and 
\cite{Harris96}. In Figure \ref{fig:MassZ Distributions} we 
show the mass-metallicity distributions of all GCs in the sample 
galaxies. To highlight the GCs that are old enough to have developed 
HBs, we divide the GCs in two subsamples:\ older or younger than 8\,Gyr 
(solid blue circles and open red squares, respectively). 
The MW GCs with extended\footnote{$\Delta V_{\rm HB}>3.5$\,mag; \cite{Lee07}.} 
HBs, i.e. EHBs, blue HBs, 
or red HBs in the bottom panel are shown with solid (blue) circles, 
(blue) crosses and open (red) squares, respectively. As Figure\ \ref{fig:MassZ
  Distributions} shows, all MW GCs more massive than
$6\times10^5M_\odot$ (indicated with a 
horizontal dashed line) have hot HBs (EHBs or BHBs), except 47\,Tuc.

As expected, the GCs in NGC\,3610 with spectroscopy are among the most massive 
ones ($M>6\times10^5M_\odot$). However, among those, only W\,28 is formally
older than 8\,Gyr with $t=11.047\pm^{6.541}_{5.054}$\,Gyr, and the age of W\,3 is
$t=7.9^{+0.8}_{-2.2}$\,Gyr, equal to 8 Gyr within the 
measurement errors. This renders them good candidates for GCs that host BHB
stars. This further supports the interpretation  
of the difference between photometric and spectroscopic ages due 
to the effect of the BHB in these metal-poor GCs as discussed in 
Section\,\ref{Subsect:Photometric Age Z}. On the other hand, the 
younger W\,31 and W\,6 ($t=6.3^{+1.3}_{-1.0}$\,Gyr, [Z/H]$=-0.498\pm^{0.082}_{0.142}$ 
and $t=5.9^{+1.8}_{-0.6}$\,Gyr, [Z/H]$=-0.610\pm^{0.169}_{0.078}$)
likely have not yet developed BHBs, which would explain the consistent 
photometric and spectroscopic ages seen in Figure\,\ref{fig:ageZ_difference}.

\section{Comparing properties of globular clusters with those of their host galaxies}\label{Sect:galaxy vs GCs properties} 

Our aim is to assess information on the star formation histories (SFHs)
of the galaxies in our sample as determined from the properties of their GCs,
and compare it to known population properties of the host galaxies themselves. This
will not only test for consistency between the two approaches, it will also
highlight the particular insights gained from the combination of deep optical and
near-IR photometry of GCs in nearby galaxies. 

\subsection{Ages and metallicities}\label{Sect:GC vs galaxies Ages and Z}

Under the assumption that massive GCs were formed during major star
formation epochs in the galaxies' assembly history, one would expect a
relation between the peak ages (and metallicities) for the GCs and the
(luminosity-weighted) ages and metallicities of their host galaxies.   
To test for such a relation, we collected recent age and metallicity
measurements for the integrated light of the sample galaxies from the literature. 
These are listed in Table\,\ref{Table:Galaxies - GCs Age Z}, and 
were derived from $B-V$
and $B-K$ colours \citep{Li07} as well as Lick indices from long-slit 
spectroscopy by \cite{Howell05} and \cite{Sanchez06} using the \cite{Thomas03} 
and Vazdekis/MILES SSP models, respectively. \cite{Sanchez06} provide 
\begin{table*}
\caption[]{Literature values of the age and metallicity of the galaxies 
in our sample derived from $B-V,B-K$ colours \citep{Li07} (columns 
2 and 3), and from integrated light spectroscopy based on Lick indices 
by \cite{Howell05} based on \cite{Thomas03} SSPs (columns 4 and 5) and 
by \cite{Sanchez06} using Vazdekis/MILES SSPs (columns 6 and 7). In 
columns 8 and 9 we show the peak age and metallicity of the metal-rich 
GC subpopulation.
}\label{Table:Galaxies - GCs Age Z}
\footnotesize
\begin{tabular*}{1\textwidth}{l|cccccc|cc}
\hline\hline
Galaxy & Age [Gyr] & [Z/H] & Age & [Z/H] & Age & [Z/H] & GCs Age & [Z/H] \\[.1cm]
& \multicolumn{2}{c}{from \citep{Li07}} & \multicolumn{2}{c}{\citep{Howell05}} & \multicolumn{2}{c}{\citep{Sanchez06}} & \multicolumn{2}{c}{from this study} \\[.1cm]
(1)&(2)&(3)&(4)&(5)&(6)&(7)&(8)&(9)\\[.1cm]
\hline
NGC\,3610	&$1.60\pm0.22$ &$0.389\pm0.034$ & $1.7\pm0.1$ &$0.76\pm0.16$ &$\cdots$ &$\cdots$ &$4\pm1$	&$0.02\pm0.15$ \\
NGC\,584	&$3.13\pm0.72$ &$0.279\pm0.060$ & $2.4\pm0.3$ &$0.61\pm0.04$ &$6.2\pm0.69$ &$0.125\pm0.031$ &$6\pm1$	&$0.19\pm0.13$ \\
NGC\,3377	&$4.08\pm1.00$ &$0.061\pm0.050$	&$3.5\pm0.8$ &$0.30\pm0.06$	&$8.9\pm1.58$	&$-0.010\pm0.068$ &$9\pm1$	&$-0.28\pm0.16$\\[.1cm]
\hline\hline
\end{tabular*}
\end{table*}
ages and metallicities from four different index-index diagrams 
([MgbFe]--H$_\beta$, Fe4383--H$_\beta$, Mgb--H$_\beta$ and CN$_2$--H$_\beta$) 
and the average of their "best nine spectral synthesis model fits" 
(giving the oldest age). We selected the average value of all these five
measurements, adding errors in quadrature.

For NGC\,3377 there are also age and metallicity estimates 
from the SAURON project \citep{Kuntschner10} at R$_e$, which are 
8.1$^{+0.4}_{-1}$\,Gyr and [Z/H]$=-0.23\pm0.04$\,dex. These were derived 
from comparing Mgb, Fe50, Fe52, and H$_\beta$ Lick indices to the Vazdekis/MILES 
SSP models. These values are similar to 
the results of \cite{Sanchez06} for NGC\,3377. The 
SAURON values are not included in Table\,\ref{Table:Galaxies - 
GCs Age Z} because the other two galaxies were not in the SAURON 
survey. An interesting result on NGC\,3377 from the latter survey is that
its inner 20\,arcsec ($\sim0.7$\,kpc) region features a relatively young stellar population
with high metallicity located in a rotating structure with disc-like
kinematics \citep{Kuntschner10}.  
This further supports the accretion/interaction-driven formation history of
NGC\,3377. 

\begin{figure}
\epsfig{file=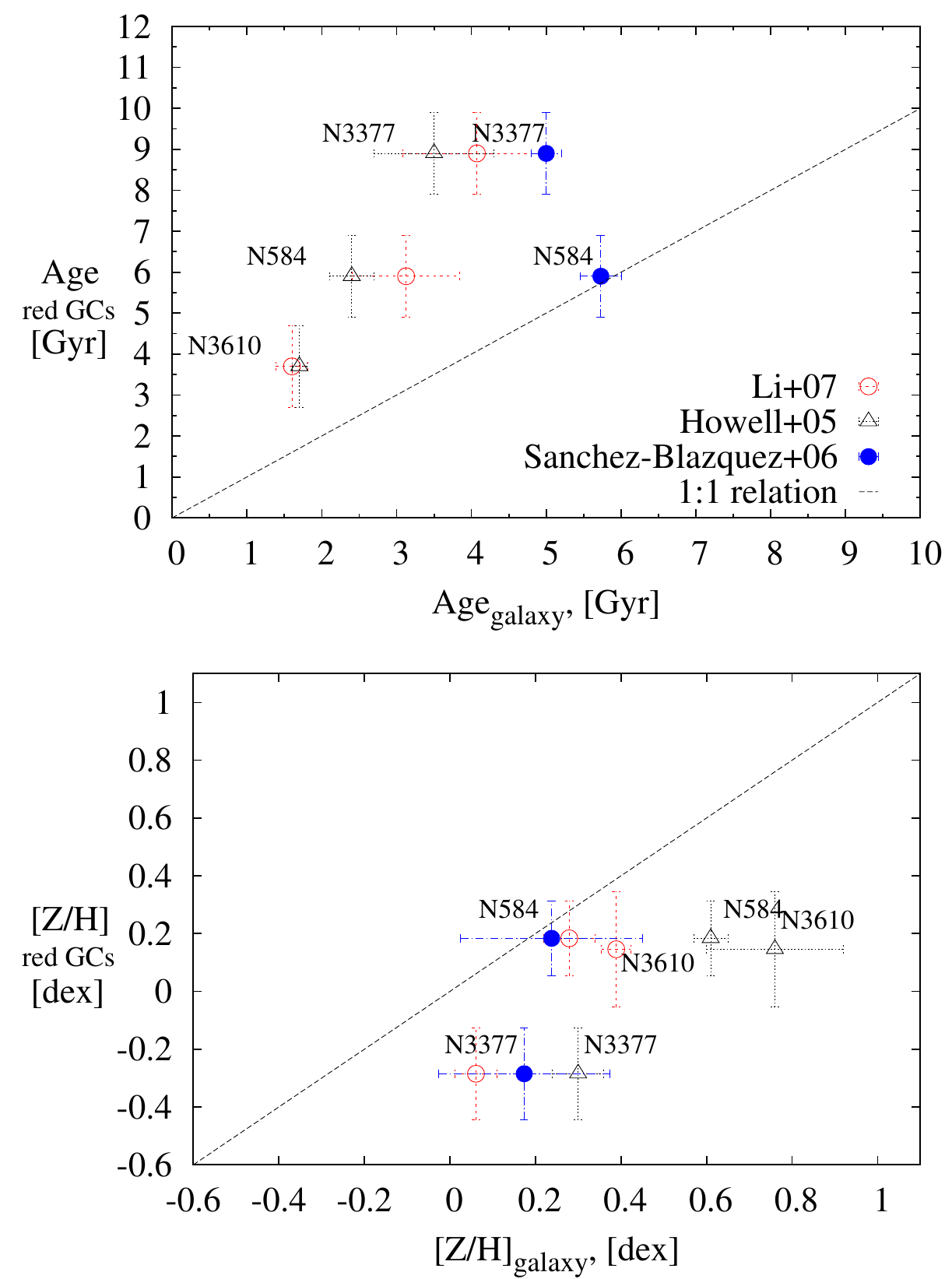, width=0.5\textwidth, bb=30 10 382 473}
\caption{Correlation between ages (top panel) and metallicities (bottom panel)
  of red globular clusters and integrated light of their host
  galaxies. Different symbol types represent different literature sources for the
  (luminosity-weighted) mean age and metallicity of the galaxy, as indicated in
  the legend. Dashed line is the one-to-one relation.
}\label{fig:Galaxies - GCs Age Z}
\end{figure}
Probing for a correlation between the luminosity-weighted stellar age and 
metallicity of the galaxy and its GCs, we plot the peak age and metallicity of
the GCs versus host galaxy age and metallicity in 
Figure\,\ref{fig:Galaxies - GCs Age Z}. It is clear that the peak age of the red  
(i.e. metal-rich) GCs in all galaxies is indeed correlated with the
luminosity-weighted mean ages and metallicity of the host  
galaxy. 

The upper panel of Figure~\ref{fig:Galaxies - GCs Age Z} suggests that the
(luminosity-weighted) ages of the sample galaxies are typically younger than
the mean ages of their bright metal-rich GCs. While this difference is likely
due in part to systematic effects related to the use of different SSP models,
this suggests, at face value, that the galaxies had a prolonged star formation
history since the major epoch(s) of massive GC formation. One scenario that
could cause this effect is that the star formation occurring more recently
than the era indicated by the peak age of the metal-rich GCs featured lower
star formation rates. This would form fewer high-mass GCs and thus cause
fewer GC detections at a given detection threshold, while still causing higher
surface brightness for the integrated galaxy light, thus yielding younger ages for
the latter. This scenario can be tested directly by probing less massive red
GCs with deeper near-IR imaging data in the future, e.g. using the NIRCAM
instrument on the James Webb Space Telescope.  

\subsection{Reconstructing galaxy SFH through its most massive GC}\label{Sect:Gal_SFR}

Following recent development on the relation between galaxy star 
formation rate (SFR) and the luminosity of its most massive cluster 
at a given epoch, we attempt to derive the galaxy star formation history (SFH)
in this section.

The formation efficiency of young massive clusters that can form in a 
galaxy has been shown to scale with the galaxy SFR \citep{Larsen&Richtler00}. 
Since the SFR is higher for higher gas densities \citep{Kennicutt98}, 
the formation of massive clusters becomes more likely as well \citep{Larsen&Richtler00,
Elmegreen&Efremov97}. However, the formation of a massive cluster can also 
be subject to statistical fluctuation resulting from a size-of-sample 
effect in which higher SFR galaxies will form more GCs and thus sample 
better the high-mass end of the cluster mass function
\citep{Billett02,Larsen02}. 
Under the assumption that the observed relation between the brightest cluster 
and the galaxy SFR \citep{Larsen02} is explained by a physical limit 
for the possible range of cluster masses regulated by the galaxy SFR, 
\cite{Weidner04} derived relations between the galaxy SFR and the initial 
(embedded) mass of its most luminous cluster, which allows one to recover 
the galaxy SFR. To obtain correctly the galaxy SFH, the present-day 
cluster mass has to be corrected for stellar evolution and dynamical 
mass-loss to obtain the mass of the cluster at the time of its formation 
\citep{Maschberger&Kroupa07}. However, the most luminous cluster in 
a galaxy is not always the most massive one. 
For starburst galaxies, the most massive cluster 
is typically young as shown in Monte Carlo simulations by \cite{Bastian08}. 
Conversely, if the duration of the Monte Carlo experiment is extended to few Gyr, 
the oldest clusters are usually the brightest \citep{Gieles09}. 
Nevertheless, \cite{Bastian08} concludes from his Monte Carlo simulations 
that the observed relation between galaxy SFR and its brightest cluster 
accurately reflects the recent SFR. Thus, Bastian employs the 
$M_V^{\rm brightest}$ vs. SFR relation from \cite{Weidner04} and 
corrects for stellar evolution, 
but not for dynamical mass loss \cite[expected to be small for massive
clusters, since the tidal/dynamical mass-loss is inversely proportional to
cluster mass;][]{Baumgardt&Makino03},  
the masses of the brightest GCs associated with a given epoch for a 
sample of major merger galaxies to derive their SFHs. Among the sample 
of their galaxies is NGC\,3610 for which they used the $M_V$ magnitudes 
from \cite{Whitmore02} and derived a peak SFR of 218 $M_\odot$\,yr$^{-1}$. 
Here, we follow the same approach to assess the SFH of the galaxies 
in our sample as follows. 

\begin{figure}
\epsfig{file=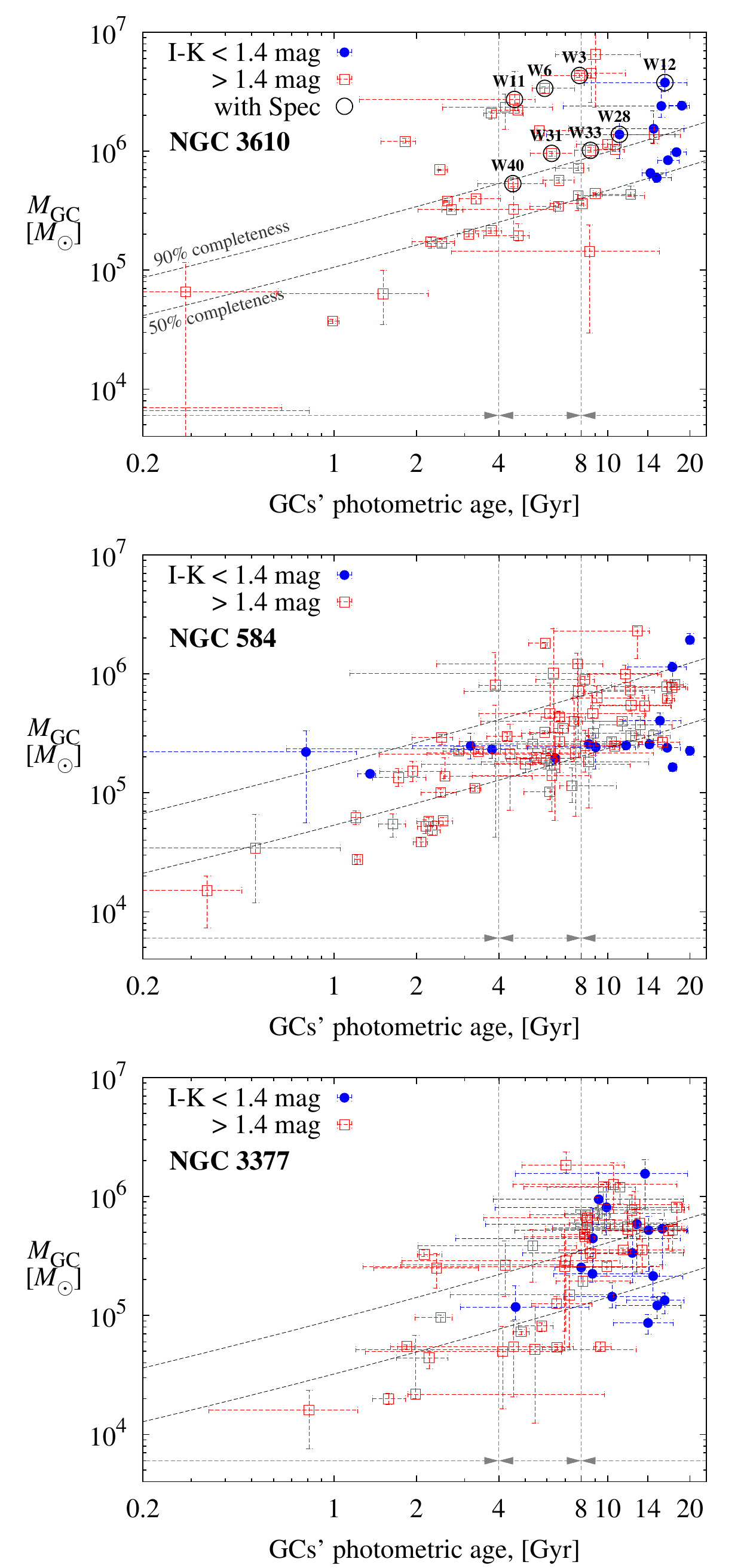, width=0.5\textwidth, bb=10 10 360 756}
\caption{Mass as a function of age for the GCs in NGC\,3610, 584 and 3377 
from top to bottom. Solid and open symbols indicate clusters bluer or 
redder than $I\!-\!K=1.4$\,mag, i.e. metal-poor or metal-rich. Vertical 
dashed lines show the age bins within which the most massive 
GCs were selected to calculate the galaxy SFRs shown in Fig.\,\ref{fig:age_GC-SFR_Gal}. 
Dashed curves show the $50\%$ and 
$90\%$ completeness levels estimated from the SSP models for all 
metallicities.
\label{fig:age-mass}
}
\end{figure}

\begin{figure}
\epsfig{file=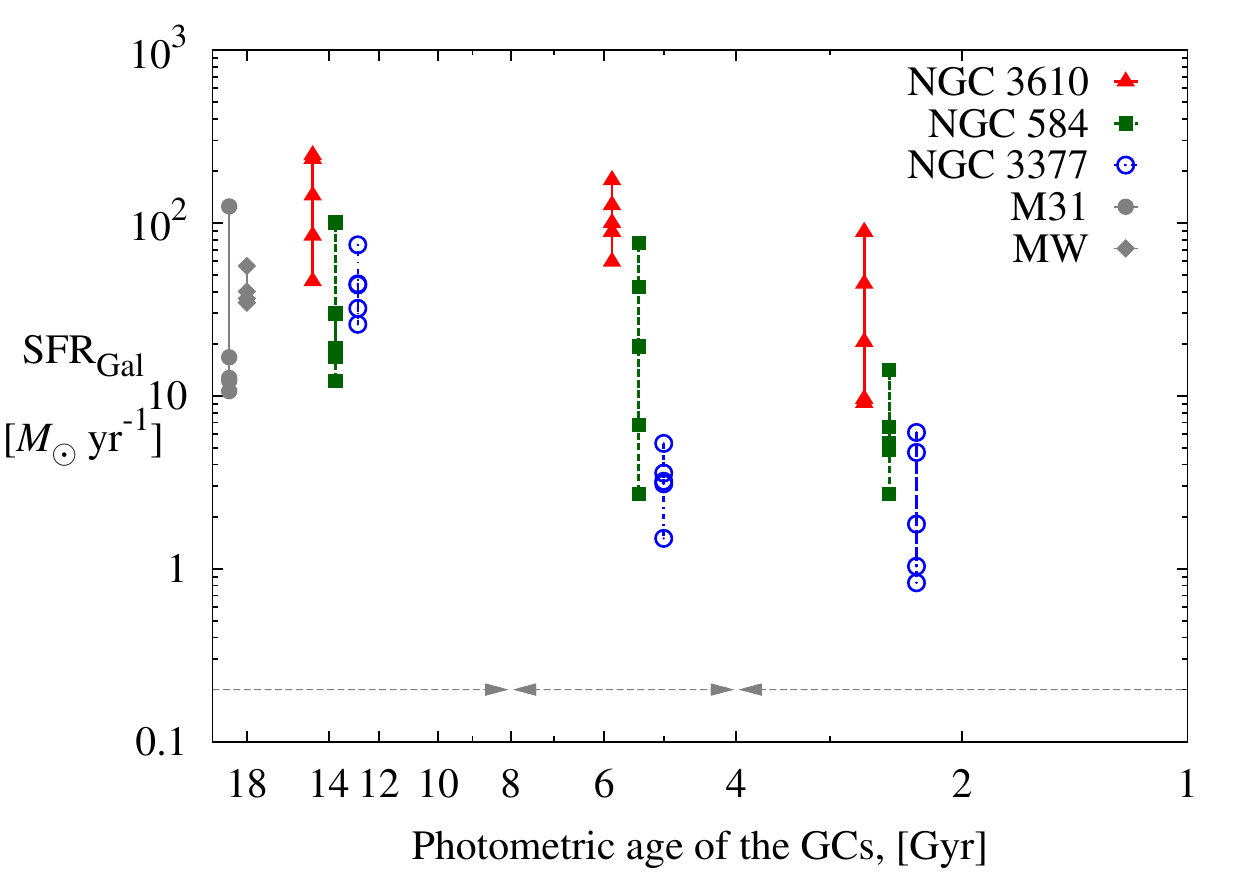, width=0.5\textwidth, bb=10 10 360 252}
\caption{Galaxy star formation rates calculated from the 
five most massive GCs (symbols connected with lines) in an age bin 
indicated with arrows at the bottom. The age positions for each galaxy 
are centered around the midpoint of the age bin, slightly offset 
from one another to improve visibility. For comparison we 
show the SFRs of the Milky Way and M\,31, calculated from their five 
most massive old GCs, which would be representative for the SFRs of 
their bulge and spheroid components.
\label{fig:age_GC-SFR_Gal}
}
\end{figure}

Using the photometric masses of our clusters, we select the five 
most massive GC in red three age bins, as shown by vertical dashed 
lines and indicated with arrows in Figure\,\ref{fig:age-mass}. 
As several studies have shown that metal-poor GCs and metal-rich GCs trace
different components of early-type galaxies, i.e., the halo and bulge/spheroidal
components, respectively \citep[e.g.,][]{Kissler-Patig97,Kundu&Whitmore98,Goudfrooij07}, we distinguish between the
metal-poor and metal-rich GCs using $I\!-\!K' = 1.4$ as a dividing colour as
before (cf.\ Section \ref{Subsect:AgeZ Distributions}). 
The mass of the most massive GCs in those age bins is then used 
to calculate the galaxy SFR using Eq.\ 7 of \citet{Weidner04}. 
To correct the mass of the GCs for mass-loss due 
to stellar evolution, we used predictions from the \cite{BC03} SSP 
models. We used the SSP $M/L_V$ at 10\,Myr to estimate the clusters' initial masses,
as in \cite{Bastian08}. In Figure\,\ref{fig:age_GC-SFR_Gal}  
we show the derived galaxy SFRs from the five most 
massive GC in three age bins ($t\!<\!4$\,Gyr\, $<\!t\!<\!8$\,Gyr, and $t\!>\!8$\,Gyr), 
where their age positions in the figure are centered around the 
midpoint of the respective age bin. The peak SFR we derive for 
NGC\,3610 (234\,$M_\odot/$\,yr) is consistent with the 218\,$M_\odot/$\,yr 
derived by \cite{Bastian08}, taking the difference in assumed distance 
moduli into account they adopt to NGC\,3610 from \cite{Whitmore02} 
($\Delta(m\!-\!M)\!=\!0.32$\,mag), as well as the adopted by them solar 
metallicity for the most massive GC. For the three 
sample galaxies, Figure\,\ref{fig:age_GC-SFR_Gal} 
shows that the peak star formation was reached at the epoch 
of formation of the oldest, metal-poor GCs. Since then, dissipative, 
gas-rich episodic (merger/accretion) events triggered the 
formation of the younger and more metal-rich stellar populations. 


Similarly, we calculate the 
SFRs of the Milky Way and M\,31 from their five most massive old 
($t\!>\!8$\,Gyr) GCs (cf Figs.\,\ref{fig:M31 GCs VIK AgeZ} and \ref{fig:age-mass}), which 
are representative for the SFRs at the time of assembly of their 
spheroid components (bulge and halo). We believe this yields a useful 
depiction of the uncertainties involved with the use of the most massive 
GC as a probe of the SFRs within these age bins. 
We do not 
expect observational bias effects in the SFR-values due to 
incompleteness (all measured "brightest" GCs are above the 90\% 
completeness level, see Fig.\,\ref{fig:age-mass}), however we can 
not exclude the possibility of missing a massive GC which might be 
located outside of the NIRI and HST WFPC2 or ACS fields 
(cf. Fig.\,\ref{fig:N3610_N584_N3377_mos2}).  Effects on the SFRs due 
to statistical fluctuations in GC mass distribution 
\citep{Maschberger&Kroupa07, Kruijssen&Cooper11} must be also kept in mind. 

The result of our analysis supports the application of the most 
massive\,GCs\,being a good tracer of the galaxy SFH.

\section{Conclusions}

We use new Gemini/NIRI $K^\prime$-band
imaging in conjunction with existing optical $V,I$ HST/WFPC2 and ACS
photometry of bright GCs in NGC\,3610, NGC\,584 and NGC\,3377 three
early-type galaxies that feature signs of recent galaxy interactions. 
The addition of the $I\!-\!K$ colour effectively breaks the age-metallicity
degeneracy present in the $V\!-\!I$ colour. 
We interpolate between \cite{BC03} SSP model tracks to derive photometric
ages, metallicities and masses 
for all GCs in the sample galaxies. For the massive GCs (${\cal{M}}\!\geq\!6\!\times\!10^5M_\odot$) 
in NGC\,3610 with available spec\-troscopic Lick indices measurements from 
the literature \citep{Strader03,Strader04}, we find that the photometric 
ages are older by $\sim\!2$\,Gyr.\,This~age~difference~is~larger~for~the~metal-poor GCs, while the photometric and spectroscopic metallicities 
are in excellent agreement (Sect.\,\ref{Subsect:Photometric Age Z}). 
We argue that this is most likely due to the presence 
of hot blue HB stars: A comparison with HB properties of Galactic GCs 
shows that {\it all} Galactic GCs with ${\cal{M}}\!\geq\!6\!\times\!10^5M_\odot$ have blue HBs,
except the metal-rich GC 47\,Tuc.  
This suggests that all NGC\,3610 GCs with spectroscopy found to be older than
$\sim$\,8 Gyr are most likely to possess a blue HB as well as the majority of
metal-poor GCs with ${\cal{M}}\!\gtrsim\!6\!\times\!10^5M_\odot$ (in {\it any} galaxy). 

We show that the peak value of the age and metallicity distributions  
of the sample galaxies GCs is correlated with the luminosity-weighted 
mean age and metallicity of the host galaxy. The GCs' age and $Z$ distributions 
are broad (Sect.\,\ref{Subsect:AgeZ Distributions}), indicating a 
prolonged star and cluster formation history. We reconstruct the galaxies' 
SFHs from the GC age distributions, using the recently observed correlation 
between the luminosity of the most massive GC at a 
given age and the galaxy SFR (Sect.\,\ref{Sect:Gal_SFR}). 

Our results support a scenario in which the star formation rate of the 
three candidate intermediate-age galaxies in our sample peaked at the epoch 
at which the oldest GCs (i.e., the metal-poor ones) formed. Subsequent
dissipative events (i.e., interactions involving gas-rich galaxies) led to the
formation of more metal-rich GCs up to a few Gyr ago. The SFHs of our 
sample galaxies as determined from their GCs suggests that the SFR was 
highest at the earliest times ($\ga$\,10 Gyr ago in all three cases. The 
latter is in large part due to the presence of massive metal-poor GCs which 
are likely associated with the early build-up of the galaxy halos. The 
strengths of subsequent star formation events occurring over the last 
$\sim$\,10 Gyr varied significantly from one galaxy to another. Among the 
three galaxies in our sample, NGC\,3610 showed the strongest SFR in all 
age bins, especially at the youngest age bin ($<$4\,Gyr ago). This is 
consistent with NGC\,3610 being the most luminous galaxy in our sample 
as well as having the youngest spectroscopically determined age.


The ability to obtain such direct insights on the SFHs of galaxies renders 
the study of galactic GC systems in the optical vs. near-IR a promising tool 
to trace galaxy formation, merging and star formation history.

\section*{Acknowledgments}

We would like to acknowledge the useful comments by the referee. IG would 
like to thank Dr. A. Kundu for kindly providing their tables with the 
HST/WFPC2 photometry of GCs in NGC\,3377 and NGC\,584.
IG is thankful~for the received support by the German\,Research\,Foundation 
(Deutsche Forschungsgemeinschaft, DFG) grant DFG\,-~Projekt BO-779/32-1. 
IYG and PG thank the Director of STScI for funding part of this research 
through a director's discretionary research grant. The authors would 
like to acknowledge useful discussions with Prof.\,P.\,Kroupa. 
This paper is based on observations obtained at the Gemini Observatory, which
is operated by the Association of Universities for Research in Astronomy,
Inc., under a cooperative agreement with the NSF on behalf of the Gemini
partnership: the National Science Foundation (United States), the Science and
Technology Facilities Council (United Kingdom), the National Research Council
(Canada), CONICYT (Chile), the Australian Research Council (Australia),
Minist\'erio da Ci\^encia e Tecnologia (Brazil) and Ministerio de Ciencia,
Tecnolog\'{\i}a e Innovaci\'on Productiva (Argentina).\\
We acknowledge the usage of the HyperLeda database \protect\href{http://leda.univ-lyon1.fr}
{http://leda.univ-lyon1.fr} \citep{Paturel03}. This research has made use 
of the NASA/IPAC Extragalactic Database (NED) which is operated by the Jet 
Propulsion Laboratory, California Institute of Technology, under contract 
with the National Aeronautics and Space Administration.

{\it Facilities:} Gemini (NIRI), HST (ACS, WFPC2).

\bibliographystyle{mn2e}
\bibliography{../../references}


\end{document}